\documentclass[aps,pra,showpacs,amssymb,nofootinbib,superscriptaddress,twocolumn,longbibliography]{revtex4-1}

\usepackage{comment}

\usepackage[usenames, dvipsnames]{color}
\usepackage{tikz}
\usetikzlibrary{shapes,backgrounds,fit,decorations.pathreplacing}
\usepackage[T1]{fontenc}
\usepackage[utf8]{inputenc}
\usepackage[croatian]{babel}
\usepackage{bbm, bm, amsbsy, amsthm, amsfonts, amsmath, dsfont, graphicx, epsfig, epstopdf, dsfont, multibib, color,relsize}
\usepackage{multirow}

\usepackage[colorlinks,breaklinks]{hyperref}
\makeatletter
\newcommand\org@hypertarget{}
\let\org@hypertarget\hypertarget
\renewcommand\hypertarget[2]{%
  \Hy@raisedlink{\org@hypertarget{#1}{}}#2%
  }
\makeatother
\usepackage[figure,table]{hypcap}
\usepackage{MnSymbol}
\usepackage{enumerate}
\usepackage{float}
\hypersetup{
	bookmarksnumbered,
	pdfstartview={FitH},
	citecolor={darkgreen},
	linkcolor={darkred},
	urlcolor={darkblue},
	pdfpagemode={UseOutlines}}
\definecolor{darkgreen}{RGB}{50,190,50}
\definecolor{darkblue}{RGB}{0,0,190}
\definecolor{darkred}{RGB}{238,0,0}
\definecolor{quantum}{RGB}{83,37,127}
\definecolor{quantumlight}{RGB}{169,146,191}
\usepackage{soul}
\newcommand{\pr}{^{\prime}}
\newcommand{\prpr}{^{\prime\prime}}
\newcommand{\ket}[1]{\ensuremath{\left|\right.\!{#1}\!\left.\right\rangle}}

\newcommand{\ketbra}[2]{\ensuremath{|{#1}\rangle\langle{#2}|}}

\newcommand{\da}[0]{\dagger}

\newcommand{\nl}{\ensuremath{\hspace*{-0.5pt}}}

\newcommand{\sub}[3]{\ensuremath{_{\hspace{#1 pt}\protect\raisebox{#2 pt}{\scriptsize{$ #3$}}}}}
\newcommand{\subtiny}[3]{\ensuremath{_{\hspace{#1 pt}\protect\raisebox{#2 pt}{\tiny{$ #3$}}}}}
\newcommand{\suptiny}[3]{\ensuremath{^{\hspace{#1 pt}\protect\raisebox{#2 pt}{\tiny{$ #3$}}}}}

\def\phi{\varphi}

\def\be{\begin{equation}}
\def\ee{\end{equation}}

\usetikzlibrary{arrows, decorations.markings}

\tikzstyle{vecArrow} = [thick, decoration={markings,mark=at position
   1 with {\arrow[semithick]{open triangle 60}}},
   double distance=1.4pt, shorten >= 5.5pt,
   preaction = {decorate},
   postaction = {draw,line width=1.4pt, white,shorten >= 4.5pt}]
\tikzstyle{innerWhite} = [semithick, white,line width=1.4pt, shorten >= 4.5pt]

\newcommand{\f}[1]{b\subtiny{-1}{-1.5}{#1}}
\newcommand{\fdag}[1]{b\subtiny{-1}{-0.5}{#1}^{\da}}

\newcommand{\tr}{\mathrm{tr}}

\newcommand{\fbra}[1]{\ensuremath{\left\langle\!\left\langle\right.\right.\! #1 \!\left.\left.\right|\hspace*{-0.75pt}\right|}}
\newcommand{\fket}[1]{\ensuremath{\left|\hspace*{-0.75pt}\left|\right.\right.\! #1 \!\left.\left.\right\rangle\!\right\rangle}}

\newcommand{\anticomm}[2]{\ensuremath{\left\{\right.\! #1 \,, #2 \!\left.\right\}}}

\newcommand{\djj}{d\kern-0.4em\char"16\kern-0.1em}
\makeatletter
\renewcommand{\p@subsection}{}
\renewcommand{\p@subsubsection}{}
\makeatother

\usepackage[customcolors]{hf-tikz}
\tikzset{style green/.style={
    set fill color=green!50!lime!60,
    set border color=white,
  },
  style cyan/.style={
    set fill color=cyan!90!blue!60,
    set border color=white,
  },
  style orange/.style={
    set fill color=orange!80!red!60,
    set border color=white,
  },
  style hordash/.style={
    set fill color=white,
    set border color=black,
  },
  hor/.style={
    above left offset={-0.09,0.25},
    below right offset={0.09,-0.05},
    #1
  },
  ver/.style={
    above left offset={-0.09,0.35},
    below right offset={0.09,-0.1},
    #1
  }
}

\usepackage[framemethod=tikz]{mdframed}
\definecolor{mycolor}{rgb}{0.122, 0.435, 0.698}
\newmdenv[innerlinewidth=0.5pt, roundcorner=4pt,linecolor=mycolor,innerleftmargin=6pt,
innerrightmargin=6pt,innertopmargin=6pt,innerbottommargin=6pt]{mybox}
\usepackage{paracol}
\usepackage{tcolorbox}
\tcbuselibrary{theorems}

\newtcbtheorem{Definitions}{Definition}%
{colback=mycolor!5,colframe=mycolor,fonttitle=\bfseries,
left=4pt,
right=4pt,
top=5pt,
bottom=5pt}{defi}

\newtcbtheorem{Lemmas}{Lemma}%
{colback=green!5,colframe=green!50!blue,fonttitle=\bfseries,
left=4pt,
right=4pt,
top=5pt,
bottom=5pt
}{lemma}

\newtcbtheorem{Question}{Question}%
{colback=magenta!5,colframe=purple!50!blue,fonttitle=\bfseries,
left=4pt,
right=4pt,
top=5pt,
bottom=5pt
}{question}

\newtcolorbox[blend into=figures]{boxdefi}[3][]
{ float*=ht,width=\textwidth,lower separated=false, center upper,
title={#2},label= def:#3,#1}

\begin{document}

\title{Teleporting Quantum Information Encoded in Fermionic Modes}
\author{Tiago Debarba}
\affiliation{Departamento Acad{\^ e}mico de Ci{\^ e}ncias da Natureza, Universidade Tecnol{\'o}gica Federal do Paran{\'a} (UTFPR), Campus Corn{\'e}lio Proc{\'o}pio, Avenida Alberto Carazzai 1640, Corn{\'e}lio Proc{\'o}pio, Paran{\'a} 86300-000, Brazil}
\author{Fernando Iemini}
\affiliation{Instituto de F\'isica, Universidade Federal Fluminense, 24210-346 Niter\'oi, Brazil}
\affiliation{ICTP, Strada Costiera 11, 34151 Trieste, Italy}
\author{Geza Giedke}
\affiliation{Donostia International Physics Center, Paseo Manuel de Lardizabal 4, E-20018 San Sebasti{\'a}n, Spain}
\affiliation{Ikerbasque Foundation for Science, Maria Diaz de Haro 3, E-48013 Bilbao, Spain}
\author{Nicolai Friis}
\email{nicolai.friis@univie.ac.at}
\affiliation{Institute for Quantum Optics and Quantum Information - IQOQI Vienna, Austrian Academy of Sciences, Boltzmanngasse 3, 1090 Vienna, Austria}

\begin{abstract}
Quantum teleportation is considered a basic primitive in many quantum information processing tasks and has been experimentally confirmed in various photonic and matter-based setups. Here, we consider teleportation of quantum information encoded in modes of a fermionic field. In fermionic systems, superselection rules lead to a more differentiated picture of entanglement and teleportation. In particular, one is forced to distinguish between single-mode entanglement swapping, and qubit teleportation with or without authentication via Bell inequality violation, as we discuss here in detail. We focus on systems subject to parity superselection where the particle number is not fixed, and contrast them with systems constrained by particle number superselection which are relevant for possible practical implementations. Finally, we analyze the consequences for the operational interpretation of fermionic mode entanglement and examine the usefulness of so-called mixed maximally entangled states for teleportation.
\end{abstract}

\date{7 May 2020}

\maketitle
\renewcommand{\figurename}{Figure}
\renewcommand{\tablename}{Table}


\section{Introduction}
\vspace*{-2mm}

Quantum teleportation refers to the transference of quantum information encoded in the complex amplitudes of an unknown quantum state of a localized system to a remote system solely via initially  entangled, local operations, and exchange of classical information. First proposed in~\cite{BennettEtAl1993}, quantum teleportation was experimentally confirmed in~\cite{Bouwmeester-etal1997}, followed soon thereafter by further experiments refining various aspects of teleportation using photon polarization~\cite{BoschiBrancaDeMartiniHardyPopescu1998, Kim-etal2001}, optical coherence~\cite{Furusawa-etal1998}, and nuclear magnetic resonance~\cite{NielsenKnillLaflamme1998}. Since then, teleportation has become a conceptual cornerstone of many tasks in quantum communication and quantum information processing. Among other methods~\cite{FriisVitaglianoMalikHuber2019}, teleportation can be seen as a way of detecting and certifying the usefulness of entanglement, because the latter is necessary to achieve a nontrivial teleportation fidelity. Practical teleportation protocols have been developed for photonic degrees of freedom, e.g., in the context of long-distance high-fidelity communication~\cite{PanDaniellGasparoniWeihsZeilinger2001, Marcikic-Gisin-etal2003, Ursin-etal2004, Ma-Zeilinger-tele2012}, chip-to-chip teleportation with applications to integrated photonic quantum technologies~\cite{Llewellyn-Thompson2020}, or multi-party settings~\cite{ZhaoChenZhangYangBriegelPan2004} relevant, e.g., for measurement-based quantum computation~\cite{RaussendorfBriegel2001, BriegelBrowneDuerRaussendorfVanDenNest2009}. In parallel to advances in photonic setups, much progress has been made for teleportation in matter-based systems~\cite{Riebe-etal2004, Olmschenk-Monroe-etal2009, NoellekeNeuznerNeuznerHahnRempeRitter2013, PfaffEtAl2014}.

Information carriers that are typically used for quantum information processing in solid-state and atomic systems (for instance, quantum dots~\cite{HansonKouwenhovenPettaTaruchaVandersypen2007}, and ions in radio-frequency traps~\cite{LeibfriedBlattMonroeWineland2003} or optical lattices~\cite{Schaetz2017}) are electrons, i.e., fermions or even the more elusive Majorana fermions~\cite{Kitaev2001, NayakSimonSternFreedmanSarma2008, SarmaFreedmanNayak2015, LutchynEtAl2018, LaflammeBaranovZollerKraus2014}. Fermionic systems are described by anticommuting operators and are subject to superselection rules~\cite{Friis2016a, AmosovFilippov2017, CabanPodlaskiRembielinskiSmolinskiWalczak2005, BanulsCiracWolf2007}, both of which require a careful approach to questions concerning correlations and entanglement~\cite{SchliemannCiracKusLewensteinLoss2001, Zanardi2002, GhirardiMarinatto2004, BanulsCiracWolf2007, ChenDjokovicGrasslZeng2013,Iemini2013, iemini14, DArianoManessiPerinottiTosini2014b, GigenaRossignoli2015, BenattiFloreaniniMarzolino2014,Lourenco19}, in particular, with regards to definition of mode subsystems~\cite{FriisLeeBruschi2013, BalachandranEtAl2013}. Nonetheless, many features known from bosonic quantum optics settings can be successfully carried over to fermions, for instance, phase-space methods for the description of Gaussian states and channels~\cite{BoteroReznik2004, Bravyi2005, CorneyDrummond2006, EislerZimboras2015, GreplovaGiedke2018, SpeeSchwaigerGiedkeKraus2018, OnumaKaluGrimmerMannMartinMartinez2019, HebenstreitJozsaKrausStrelchukYoganathan2019}. Research in this direction has previously mostly been confined to the domain of theoretical analysis, but impressive technological advances in the control and manipulation of individual electrons~\cite{McNeilEtAl2011, HermelinEtAl2011, BertrandEtAl2016, Ford2017, FujitaEtAl2017}, as well as in the generation of electronic mode-entangled states~\cite{HoferDasenbrookFlindt2017} motivate further studies of fermionic entanglement also from a practical perspective.

A key open problem in this area concerns the assignment of clear operational meaning to fermionic mode entanglement and its quantifiers. That is, fermionic systems with variable or indefinite numbers of particles (but subject to superselection rules) allow for different ways of quantifying entanglement, see, e.g.,~\cite{CabanPodlaskiRembielinskiSmolinskiWalczak2005, BanulsCiracWolf2007}. But what do these quantifiers tell us about the usefulness of the corresponding states in practical tasks? Consider a single fermionic excitation in an equally weighted superposition of two different field modes in analogy to a single photon that is delocalized in a two-path interferometer. Formally, such a state can be seen as being maximally entangled: The state of either mode is maximally mixed. But is this type of entanglement operationally meaningful? For instance, can one use it to violate a Bell inequality? If a quantum state allows such a violation then it can be unambiguously concluded that the state is entangled. For fermionic mode entanglement it was shown in~\cite{DasenbrookBowlesBohrBraskHoferFlindtBrunner2016} that this is indeed possible provided that two locally processed copies of a maximally entangled two-mode state (four modes in total) are used. While this can be seen as a device-independent certification of the entanglement of the state, one may nonetheless wonder to what (further) practical use this fermionic mode entanglement can be put. Here, we therefore investigate teleportation using fermionic mode entanglement as a resource.

Inspired by initial work in this direction~\cite{MorgenshternReznikZalzberg2008}, we review and more closely examine the different ways of interpreting the standard teleportation protocol~\cite{BennettEtAl1993} for the task of teleporting fermionic quantum information in the presence of superselection rules (SSRs). Previous proposals and experiments~\cite{SamuelssonSukhorukovBuettiker2004, BeenakkerKindermann2004, RiebeMonzKimVillarSchindlerChwallaHennrichBlatt2008, SteffenEtAl2013, QiaoEtAl2019} leave no doubt that teleportation using fermionic quantum systems is indeed possible. However, here we aim to identify the minimal resources for fermionic teleportation in order to better understand the operational meaning of the fundamental unit of fermionic mode entanglement. In particular, we are interested in the subtle consequences that SSRs imply for the usefulness of fermionic entangled states as resources for teleportation. As we discuss, the parity superselection rule (P-SSR) imposes restrictions that require a more differentiated specification of what is meant by \emph{`teleporting quantum information'} in the first place.

For a single fermionic mode that is not entangled with any other mode(s), the parity SSR implies that the encoded information is classical. Consequently, teleporting such a state requires no shared entanglement in principle. However, when the mode in question is entangled with an another (auxiliary) mode, an entangled resource state is necessary for teleportation-based entanglement swapping. When more than one mode is considered, the equivalent of one qubit of quantum information can be directly encoded in the teleported state (e.g., dual-rail encoding in two modes). However, we find that the corresponding protocols require more resources as compared to standard qubit teleportation. To transfer the complex amplitudes of a single qubit, one can make do with sharing a single maximally entangled fermionic mode pair and two bits of classical information (a fermionic single-mode teleportation channel), but one also needs to transfer additional information about the teleported state (the state of the second mode) via a fermionic quantum channel. This channel may be realized by another fermionic single-mode teleportation channel, increasing the required resources to two copies of maximally entangled two-mode states, along with the usual two bits of classical information.

Within the framework of these variations of standard teleportation we discuss the consequences of further restrictions. In particular, we consider the potential of fermionic Gaussian states and operations for teleportation, as well as the limitations imposed by particle number superselection, which is highly relevant for potential experimental implementations (in particular, using state-of-the-art methods in electron quantum optics~\cite{OlofssonPottsBrunnerSamuelsson2020}). Finally, we apply our findings to understand the wider implications for the quantification of fermionic entanglement, especially with a view to the notion of `mixed maximally entangled' (MME) fermionic states~\cite{DArianoManessiPerinottiTosini2014a} and their usefulness for teleportation.

The paper is structured as follows. In Sec.~\ref{sec:framework}, we briefly discuss the mathematical framework of fermionic modes and their entanglement. In Sec.~\ref{sec:Fermionic teleportation}, we then turn to teleportation. First, we review the standard protocol for qubit teleportation as a backdrop and discuss how fermionic teleportation deviates from this well-established paradigm in Sec.~\ref{sec:fermions vs qubits}. We then analyze protocols for teleporting the state of a single fermionic mode in Sec.~\ref{sec:scenario I}, as well as their implementation via fermionic Gaussian operations in Sec.~\ref{sec:impl ferm Gauss}, before we turn to teleportation of states of several modes in Sec.~\ref{subsubsec:role of mode Apr}. In Sec.~\ref{sec:particle number SSR teleportation}, we then discuss how the presented teleportation schemes (and potential practical implementations in electron quantum optics~\cite{OlofssonPottsBrunnerSamuelsson2020}) are influenced by the additional constraint of a SSR for the particle number. The implications on the quantification of fermionic (mode) entanglement are analyzed in Sec.~\ref{sec:implications}, with a special view to MME states.

\vspace*{-2mm}
\section{Framework}\label{sec:framework}
\vspace*{-2mm}

\subsection{Fermionic modes}

We consider quantum information encoded in the modes of a fermionic field\footnote{For now, we impose no further constraints such as a particular (half-integer) spin, fixed mass or charge on the field excitations, but we discuss such restrictions in Sec.~\protect\ref{sec:particle number SSR teleportation}.}. To each mode labelled $i$ we associate a pair of fermionic mode operators $b_{i}$ and $b_{i}^{\dagger}$, which satisfy
\begin{align}
    \anticomm{b\sub{0}{-1.5}{i}}{b_{j}^{\dagger}}_{+}   &=\,\delta_{ij}\,,
    \qquad
    \anticomm{b\sub{0}{-1.5}{i}}{b\sub{0}{-1.5}{j}}_{+}   \,=\,0\,\ \ \forall i,j,
    \label{eq:anticomm relations}
\end{align}
where $\anticomm{.}{.}_{+}$ denotes the anticommutator. The corresponding Fock space is constructed by the action of the creation operators $b_{i}^{\dagger}$ on the vacuum state $\fket{\!0\!}$, which itself is annihilated by all annihilation operators~$b_{i}$, i.e., $b_{i}\fket{\!0\!}=0\ \forall\,i$. The creation operators~$b_{i}^{\dagger}$ populate the vacuum with single fermions, that is, $b_{i}^{\dagger}\fket{\!0\!}=\fket{\!1_{i}\!}$. Due to the indistinguishability of the particles the tensor product of single-particle states needs to be antisymmetrized when two or more fermions are created. Here, we use the convention
\begin{align}
    b_{k}^{\dagger}b_{k\pr}^{\dagger}\fket{\!0\!}   &=\,\fket{\!1_{k}\!}\wedge\fket{\!1_{k\pr\!}\!}=\fket{\!1_{k}\!}\fket{\!1_{k\pr\!}\!}\,,
    \label{eq:fermion two particle state}
\end{align}
where we use double-lined notation to indicate the antisymmetrized wedge product ``$\wedge$" between two or more single-mode state vectors \emph{with} particle content (in contrast to the notation $\ket{\cdot}\ket{\cdot}=\ket{\cdot}\otimes\ket{\cdot}$ for a tensor product), i.e., we have $\fket{1_{k}}\fket{1_{k\pr}}=-\fket{1_{k\pr}}\fket{1_{k}}$, whereas combinations of states with and \emph{without} particle content satisfy $\fket{0}\fket{1_{k}}=\fket{1_{k}}\fket{0}=\fket{1_{k}}$. With this definition at hand, arbitrary pure states on the Fock space can be written as
\begin{align}
    \fket{\!\Psi\!}    &=
    \gamma_{0}\fket{0}+\sum\limits_{i=1}^{n}\gamma_{i}\fket{\!1_{i}\!}+
    \sum\limits_{j,k}\gamma_{jk}\fket{\!1_{j}\!}\fket{\!1_{k}\!}+\ldots\ .
    \label{eq:fermionic Fock space general state rewritten}
\end{align}
However, the parity superselection rule (see, e.g.,~\cite{Friis2016a, AmosovFilippov2017, CabanPodlaskiRembielinskiSmolinskiWalczak2005, BanulsCiracWolf2007}) implies that coherent superpositions of even and odd numbers of fermions cannot exist. For instance, a general pure state of two modes $A$ and $A\pr$ of the form
\begin{align}
    \fket{\!\Psi\!}\subtiny{-1}{0}{AA\pr}    &=
    \gamma_{0}\fket{0}+\gamma\subtiny{-1}{0}{A}\fket{\!1\subtiny{-1}{0}{A}\!} +\gamma\subtiny{-1}{0}{A\pr}\fket{\!1\subtiny{-1}{0}{A\pr}\!}
    \nonumber\\[1mm]
    &\ \ +\gamma\subtiny{-1}{0}{AA\pr}\fket{\!1\subtiny{-1}{0}{A}\!}\fket{\!1\subtiny{-1}{0}{A\pr}\!},
    \label{eq:fermionic Fock space general state two-modes}
\end{align}
must be an even- or odd-parity state, i.e., the probability amplitudes must satisfy either
\begin{align*}
    \phantom{\text{or}}\ \ \gamma\subtiny{-1}{0}{A}=\gamma\subtiny{-1}{0}{A\pr}=0\,,   &\ \ \ |\gamma_{0}|^{2}+|\gamma\subtiny{-1}{0}{AA\pr}|^{2}=1\ \ \ &\text{(even parity)},\\[1mm]
    \text{or}\ \ \gamma_{0}=\gamma\subtiny{-1}{0}{AA\pr}=0\,,   &\ \ \ |\gamma\subtiny{-1}{0}{A}|^{2}+|\gamma\subtiny{-1}{0}{A\pr}|^{2}=1 \ \ \ &\text{(odd parity)}.
\end{align*}
While coherent superpositions of states with different parity are thus forbidden, incoherent mixtures are still possible. In particular, this implies that any fermionic single-mode state must be of the form
\begin{align}
    \rho\subtiny{0}{0}{A}    &=\,p\,\fket{0}\!\!\fbra{0}\,+\,(1-p)\,\fket{1\subtiny{-1}{0}{A}}\!\!\fbra{1\subtiny{-1}{0}{A}}\,,
    \label{eq:single mode state}
\end{align}
for $0\leq p\leq 1$.


\subsection{Entanglement of fermionic modes}\label{sec:entanglement of ferm modes}

The parity superselection rule also has interesting consequences for defining entanglement between fermionic mode subsystems. In principle, one can define subsystems containing complementary, non-overlapping sets of fermionic modes and consider (quantum) correlations between them, see, for instance~\cite{BanulsCiracWolf2007, CabanPodlaskiRembielinskiSmolinskiWalczak2005, FriisLeeBruschi2013, ChenDjokovicGrasslZeng2013, DArianoManessiPerinottiTosini2014b,debarba2017}. In particular, the `local' operators assigned to different subsystems need not commute as one would usually assume, but they can also anticommute. In fact, there is no particular reason why the modes in question need to be spatially separated. For instance, one may consider two spatially overlapping (but orthogonal) field modes with different frequencies. In any case, particular care must be taken to deal with the definition of partial traces~\cite{FriisLeeBruschi2013, BalachandranEtAl2013} to avoid ambiguities such as those discussed in~\cite{MonteroMartinMartinez2011b, BradlerJauregui2012, MonteroMartinMartinez2012a}. In other words, two-mode states like
\begin{subequations}
\label{eq:two entangled fermionic modes}
\begin{align}
    \text{even:}\ \ \fket{\psi\suptiny{0}{0}{\mathrm{e}}}\subtiny{-1}{0}{AA\pr}   &=\,\alpha\,\fket{0}\,+\,\beta\,\fket{1\subtiny{-1}{0}{A}}\fket{1\subtiny{-1}{0}{A\pr}},\label{eq:two entangled fermionic modes even}\\[1mm]
    \text{odd:}\ \ \fket{\psi\suptiny{0}{0}{\mathrm{o}}}\subtiny{-1}{0}{AA\pr}      &=\,\alpha\,\fket{1\subtiny{-1}{0}{A\pr}}\,+\,\beta\,\fket{1\subtiny{-1}{0}{A}},
    \label{eq:two entangled fermionic modes odd}
\end{align}
\end{subequations}
can be regarded as entangled (for $\alpha\beta\not=0$). In particular, we can define a basis of
\emph{maximally entangled two-mode states} $\fket{\Phi^{\pm}}\subtiny{-1}{0}{A\nl B}$ and $\fket{\Psi^{\pm}}\subtiny{-1}{0}{A\nl B}$ given by
\begin{subequations}\label{eq: maximally entangled states}
\begin{align}
    \fket{\Phi^{\pm}}\subtiny{-1}{0}{A\nl B}    &=\,
    \tfrac{1}{\sqrt{2}}\bigl(\fket{0}\pm\fket{1\subtiny{-1}{0}{A}}\fket{1\subtiny{-1}{0}{B}}\bigr),\\[1mm]
    \fket{\Psi^{\pm}}\subtiny{-1}{0}{A\nl B}    &=\,
    \tfrac{1}{\sqrt{2}}\bigl(\fket{1\subtiny{-1}{0}{B}}\pm\fket{1\subtiny{-1}{0}{A}}\bigr).
\end{align}
\end{subequations}

The parity superselection rule leads to some interesting differences with respect to the corresponding two-qubit states. First, one notes that measurements in any local single-mode basis other than the `computational' basis $\{\fket{0},\fket{1_{A}}\}$ are prevented by parity superselection. At the same time, measurements in a single product basis are not sufficient to distinguish entanglement from purely classical correlations. Consequently, two copies of each state need to be processed simultaneously to allow for the violation of a Bell inequality~\cite{DasenbrookBowlesBohrBraskHoferFlindtBrunner2016} (see also Ref.~\cite{Friis2016b}). Second, the superselection rule also restricts the physically allowed pure-state decompositions for any given mixed state, which enters in convex-roof entanglement measures. For instance, consider the entanglement of formation (EOF)~\cite{BennettDiVincenzoSmolinWootters1996}, defined as
\begin{align}\label{eq: entanglement of formation}
    \mathcal{E}_{\mathrm{oF}}(\rho) &:=\,\inf_{\mathcal{D}(\rho)}\sum\limits_{i}p_{i}\,S(\rho\subtiny{-1}{0}{A}\suptiny{0}{0}{(i)})\,,
\end{align}
where $S(\rho)=-\tr\bigl(\rho\log(\rho)\bigr)$ is the von~Neumann entropy and the infimum is taken over all pure-state decompositions, that is, $\mathcal{D}(\rho)$ is normally taken to be the set of all sets $\{(p_{i},\ket{\psi_{i}})\}_{i}$ for which $\rho=\sum_{i}p_{i}\ketbra{\psi_{i}}{\psi_{i}}$, with $\sum_{i}p_{i}=1$ and $0\leq p_{i}\leq1$. For fermionic modes it can now be argued~\cite{DArianoManessiPerinottiTosini2014a} that the set $\mathcal{D}(\rho)$ should be restricted to allow only pure state decompositions $\{(p_{i},\ket{\psi_{i}})\}_{i}$ where the states $\ket{\psi_{i}}$ satisfy the parity superselection rule. This assumption leads to the notion of mixed maximally entangled (MME) states~\cite{DArianoManessiPerinottiTosini2014a}, i.e., mixtures of maximally entangled pure states from different parity subspaces that are still as entangled (according to the value of the superselected EOF) as the individual pure states. The question that remains is, what is the operational significance of the value of the fermionic EOF?

Here, we therefore want to investigate the role of superselection rules and fermionic entanglement in teleportation protocols. More specifically, we aim to extend previous work~\cite{MorgenshternReznikZalzberg2008} in this direction and identify if and how quantum information encoded in fermionic modes can be teleported, which resources need to be shared and which information needs to be communicated, before we return to a discussion of the implications for fermionic entanglement in Sec.~\ref{sec:implications}.


\section{Fermionic teleportation}\label{sec:Fermionic teleportation}

\subsection{Fermionic versus qubit teleportation}\label{sec:fermions vs qubits}

\begin{figure}[ht!]
\begin{center}
\hspace*{6mm}
\includegraphics[width=0.3\textwidth,trim={0cm 0cm 0cm 0cm},clip]{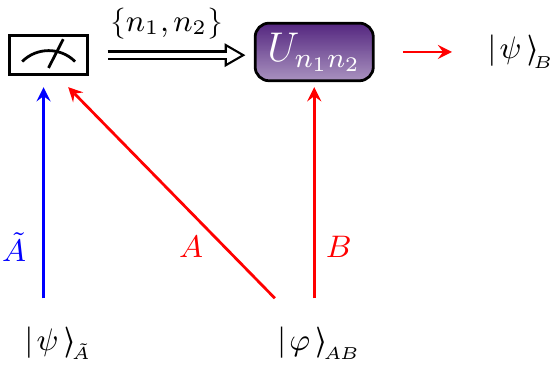}
\vspace*{-3mm}
\caption{Qubit teleportation. Teleporting one qubit of quantum information, encoded in the single-qubit state $\ket{\psi}\subtiny{-1}{0}{\tilde{A}}$ from qubit $\tilde{A}$ to qubit $B$, requires sharing one maximally entangled two-qubit state $\ket{\varphi}\subtiny{-1}{0}{A\nl B}$ (1 `ebit') and communicating two bits of classical information with bit values $n_{1}$ and $n_{2}$, respectively.}
\label{fig:qubittele}
\end{center}
\end{figure}

To set the stage for explaining fermionic teleportation scenarios, let us briefly sketch the standard protocol for teleporting a single qubit between two observers called Alice and Bob, as illustrated in Fig.~\ref{fig:qubittele}. There, to teleport an unknown state $\ket{\psi}\subtiny{-1}{0}{\tilde{A}}$ of qubit $\tilde{A}$, held by Alice, a maximally entangled two-qubit state $\ket{\Phi}\subtiny{-1}{0}{A\nl B}$ of qubits $A$ and $B$ is shared between Alice and Bob. Then, a projective measurement in a maximally entangled two-qubit basis is performed on qubits $A$ and $\tilde{A}$ by Alice. The result of the measurement, encoded in two classical bits with values $n_{1}$ and $n_{2}$, is then sent to Bob, who applies a corresponding unitary $U_{n_{1}n_{2}}$ on qubit $B$, recovering the state $\ket{\psi}\subtiny{-1}{0}{B}$.

The basic observation to understand where teleportation of fermionic quantum information deviates from standard teleportation of qubits~\cite{BennettEtAl1993} is that parity superselection implies that single-mode states of fermionic fields are of the form of Eq.~(\ref{eq:single mode state}). On the one hand, this means that a single mode can locally only encode classical information (the equivalent of a classical bit). Consequently, teleportation of the quantum information stored solely in a single fermionic mode state is trivial: One can simply measure the state, send the result as one bit of classical information via a classical channel and prepare the corresponding state at the other end. On the other hand, the entropy of the single-mode state can arise from lack of information but also from entanglement with another mode. That is, the state $\rho\subtiny{0}{0}{\tilde{A}}$ that is to be teleported and which is of the form of Eq.~(\ref{eq:single mode state}) may be the marginal of a two-mode state $\rho\subtiny{0}{0}{\tilde{A}\nl\tilde{A}\pr}$, for instance, as in Eqs.~(\ref{eq:two entangled fermionic modes even}) or~(\ref{eq:two entangled fermionic modes odd}) with $p=|\alpha|^{2}$, or even an incoherent mixture of the two. In this case, a purely classical `measure and prepare' protocol would transfer classical information stored locally in mode $\tilde{A}$, but would not be able to preserve entanglement with mode $\tilde{A}\pr$.

\begin{figure}[ht!]
\begin{center}
\includegraphics[width=0.32\textwidth,trim={0cm 0cm 0cm 0cm},clip]{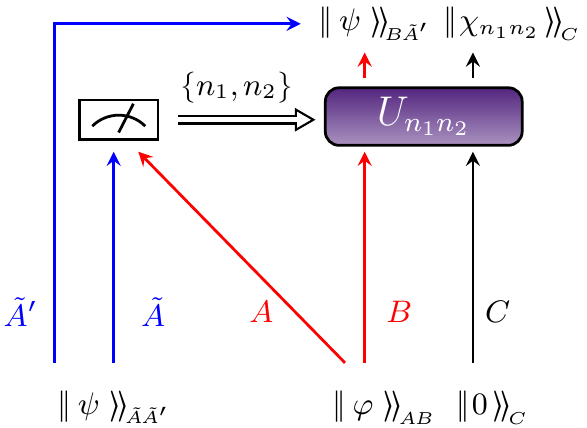}
\caption{Fermionic single-mode teleportation. In this scenario, quantum information encoded mode $\tilde{A}$ is teleported to mode $B$ by sharing one maximally entangled two-mode state $\fket{\varphi}\subtiny{-1}{0}{A\nl B}$ (1 `fbit') and communicating two bits of classical information with bit values $n_{1}$ and $n_{2}$. The four possible values of these bits correspond to the four possible measurement outcomes of a measurement in the basis $\{\fket{\Phi^{\pm}}\subtiny{-1}{0}{\tilde{A}\nl A},\fket{\Psi^{\pm}}\subtiny{-1}{0}{\tilde{A}\nl A}\}$. Finally, a unitary $U_{n_{1}n_{2}}$ that depends on the bit values $n_{1}$ and $n_{2}$ is applied to the modes $B$ and $C$ to recover the teleported state in mode $B$. Due to the parity superselection rule, some of the unitaries $U_{n_{1}n_{2}}$ require changing the state of the auxiliary mode $C$ from $\fket{0}$ to $\fket{1\subtiny{-1}{0}{C}}$. Mode $\tilde{A}$ may initially be in an entangled pure state $\fket{\psi}\subtiny{-1}{0}{\tilde{A}\nl \tilde{A}'}$ or an arbitrary mixed state $\rho\subtiny{0}{0}{\tilde{A}\nl \tilde{A}\pr}$ with mode $\tilde{A}\pr$, as we discuss in more detail in Sec.~\ref{subsubsec:role of mode Apr}. The details of the teleportation protocol do not depend on the parity sector of the state $\fket{\psi}\subtiny{-1}{0}{\tilde{A}\nl \tilde{A}'}$.}
\label{fig:telescenario1}
\end{center}
\end{figure}

To consider teleportation of quantum information using fermionic modes in any nontrivial way, we hence first have to decide what we mean by `quantum information': If by `quantum information' we mean the state of a system that itself contains only classical information but which might be entangled with another system, then we can consider teleporting the state of a single fermionic mode, as we discuss in Sec.~\ref{sec:scenario I}. If, on the other hand, we require the transfer of the equivalent of \emph{one qubit} of quantum information, then we either have to relax the rules of the teleportation protocol (see Sec.~\ref{subsubsec:role of mode Apr}) or consider the teleportation of an \emph{entangled two-mode state} from Eq.~(\ref{eq:two entangled fermionic modes}), since this definition implies that a single fermionic mode cannot contain quantum information.

As in the teleportation using qubits, the teleportation of fermionic quantum information of any kind requires two resources:
\begin{enumerate}[(i)]
    \item{Shared entangled states: For qubits, one usually considers the number of `ebits', i.e., shared maximally entangled qubit pairs. Here, we consider the number of required fermionic ebits, i.e., maximally entangled two-mode states, which we call `fbits'.}
    \item{Sending classical information (in bits).}
\end{enumerate}
For qubits, the minimal amount of resources for teleportation of 1 qubit is 1 ebit and 2 bits. For the teleportation of fermionic quantum information, the minimally required resources depend on the particular scenario one considers. In the following, we will discuss these different scenarios and the corresponding resources, advantages, and drawbacks.


\subsection{Fermionic single-mode teleportation}\label{sec:scenario I}

For the teleportation of fermionic quantum information, we
explore a situation where Alice wishes to teleport the state of a single mode labelled $\tilde{A}$ to Bob, as illustrated in Fig.~\ref{fig:telescenario1}. The mode $\tilde{A}$ may (potentially) be entangled with another mode $\tilde{A}\pr$ that is itself not necessarily teleported and whose role we discuss in more detail in Sec.~\ref{subsubsec:role of mode Apr}. For simplicity, let us for now assume that the two modes are prepared in the state $\fket{\psi}\subtiny{-1}{0}{\tilde{A}\nl \tilde{A}\pr}$ as in Eq.~(\ref{eq:two entangled fermionic modes}). We further assume that Alice and Bob share one maximally entangled fermionic two-mode state $\fket{\phi}\subtiny{-1}{0}{A\nl B}$, i.e., 1 fbit, as a resource to teleport the state of mode $\tilde{A}$ from Alice to Bob. Alice then performs a projective measurement with respect to the basis $\{\fket{\Phi^{\pm}}\subtiny{-1}{0}{\tilde{A}\nl A},\fket{\Psi^{\pm}}\subtiny{-1}{0}{\tilde{A}\nl A}\}$ on the modes $\tilde{A}$ and $A$.


\subsubsection{Even-parity resource states}

With the mentioned choice of measurement basis in mind, we can write the joint initial state $\fket{\psi}\subtiny{-1}{0}{\tilde{A}\nl \tilde{A}\pr}\fket{\phi}\subtiny{-1}{0}{A\nl B}$ for the specific case where $\fket{\psi}\subtiny{-1}{0}{\tilde{A}\nl \tilde{A}\pr}=\fket{\psi\suptiny{0}{0}{\mathrm{e}}}\subtiny{-1}{0}{\tilde{A}\nl \tilde{A}\pr}$ and $\fket{\phi}\subtiny{-1}{0}{A\nl B}=\fket{\Phi^{\pm}}\subtiny{-1}{0}{A\nl B}$, i.e., both states have even parity. Then, we have
\begin{align}
    \fket{\psi\suptiny{0}{0}{\mathrm{e}}}\subtiny{-1}{0}{\tilde{A}\nl \tilde{A}\pr}
    \fket{\Phi^{\pm}}\subtiny{-1}{0}{A\nl B}  &=\,
    \tfrac{1}{2}\Bigl[
        \fket{\Phi^{+}}\subtiny{-1}{0}{\tilde{A}\nl A}
        \bigl(\alpha\fket{0}\pm\beta\fket{1\subtiny{-1}{0}{B}}\fket{1\subtiny{-1}{0}{\tilde{A}\pr}}\bigr)
        \nonumber\\[1mm]
        &\ \ +
        \fket{\Phi^{-}}\subtiny{-1}{0}{\tilde{A}\nl A}
        \bigl(\alpha\fket{0}\mp\beta\fket{1\subtiny{-1}{0}{B}}\fket{1\subtiny{-1}{0}{\tilde{A}\pr}}\bigr)
        \nonumber\\[1mm]
        &\ \ \pm
        \fket{\Psi^{+}}\subtiny{-1}{0}{\tilde{A}\nl A}
        \bigl(\alpha\fket{1\subtiny{-1}{0}{B}}\pm\beta\fket{1\subtiny{-1}{0}{\tilde{A}\pr}}\bigr)
        \nonumber\\[1mm]
        &\ \ \pm
        \fket{\Psi^{-}}\subtiny{-1}{0}{\tilde{A}\nl A}
        \bigl(\alpha\fket{1\subtiny{-1}{0}{B}}\mp\beta\fket{1\subtiny{-1}{0}{\tilde{A}\pr}}\bigr)\Bigr].
        \label{eq:even even}
\end{align}

The measurement with respect to the basis $\{\fket{\Phi^{\pm}}\subtiny{-1}{0}{\tilde{A}\nl A},\fket{\Psi^{\pm}}\subtiny{-1}{0}{\tilde{A}\nl A}\}$ results in one of four possible outcomes corresponding to the four orthogonal basis states. Alice encodes the outcome in two classical bits, $n_{1}$ and $n_{2}$, and communicates them to Bob via a classical channel. If the outcome suggests that the modes $\tilde{A}$ and $A$ have been projected onto the state $\fket{\Phi^{\pm}}\subtiny{-1}{0}{\tilde{A}\nl A}$, i.e., the initially shared resource state, then the modes $B$ and $\tilde{A}\pr$ are left in the `correct' state $\fket{\psi\suptiny{0}{0}{\mathrm{e}}}\subtiny{-1}{0}{B\! \tilde{A}\pr}=\alpha\fket{0}+\beta\fket{1\subtiny{-1}{0}{B}}\fket{1\subtiny{-1}{0}{\tilde{A}\pr}}$, without any further action. If the obtained outcome is $\fket{\Phi^{\mp}}\subtiny{-1}{0}{\tilde{A}\nl A}$, i.e., an outcome in the same (even) parity sector as the resource state but with a relative phase of $\pi$, then a phase flip transformation is required which can be represented by the unitary
\begin{align}
    U_{\pi} &=
    \exp\bigl(i\pi b\subtiny{0}{0}{B}^{\dagger}b\subtiny{-0.5}{-1.5}{B}\bigr),
    \label{eq:Upi}
\end{align}
which maps $\fket{1\subtiny{-1}{0}{B}}$ to $-\fket{1\subtiny{-1}{0}{B}}$ and leaves all other modes invariant. When the outcome corresponds to a state in the opposite (odd) parity sector, i.e., $\fket{\Psi^{\pm}}\subtiny{-1}{0}{\tilde{A}\nl A}$, then Bob needs to apply a unitary $U\subtiny{-1}{0}{\mathrm{P}}$ to switch the parity of the state in the modes $B$ and $\tilde{A}\pr$. Due to parity superselection this is of course only possible via a parity conserving operation on a larger Hilbert space. We therefore append an auxiliary mode $C$ that is initially not populated and define the unitary $U\subtiny{-1}{0}{\mathrm{P}}$ as
\begin{align}
    U\subtiny{-1}{0}{\mathrm{P}} &=
    \bigl(\f{C}+\fdag{C}\bigr)\bigl(\f{B}-\fdag{B}\bigr),
\end{align}
\vspace*{-2mm}
\begin{table}[ht!]
\begin{center}
\renewcommand{\arraystretch}{1.7}
\begin{tabular}{ |c||c|c|c|c| }
\hline
 &  $\fket{\Phi^{+}}\subtiny{-1}{0}{A\nl B}$ &  $\fket{\Phi^{-}}\subtiny{-1}{0}{A\nl B}$ &  $\fket{\Psi^{+}}\subtiny{-1}{0}{A\nl B}$ &  $\fket{\Psi^{-}}\subtiny{-1}{0}{A\nl B}$\\
\hline\hline
$\fket{\Phi^{+}}\subtiny{-1}{0}{\tilde{A}\nl A}$ & $\mathds{1}$ & $U_{\pi}$ & $U\subtiny{-1}{0}{\mathrm{P}}$ & $U\subtiny{-1}{0}{\mathrm{P}}U_{\pi}$\\
\hline
$\fket{\Phi^{-}}\subtiny{-1}{0}{\tilde{A}\nl A}$  & $U_{\pi}$ & $\mathds{1}$ & $U\subtiny{-1}{0}{\mathrm{P}}U_{\pi}$ & $U\subtiny{-1}{0}{\mathrm{P}}$ \\
\hline
$\fket{\Psi^{+}}\subtiny{-1}{0}{\tilde{A}\nl A}$  & $U\subtiny{-1}{0}{\mathrm{P}}U_{\pi}$ & $U\subtiny{-1}{0}{\mathrm{P}}$ & $U_{\pi}$ & $\mathds{1}$ \\
\hline
$\fket{\Psi^{-}}\subtiny{-1}{0}{\tilde{A}\nl A}$ & $U\subtiny{-1}{0}{\mathrm{P}}$ & $U\subtiny{-1}{0}{\mathrm{P}}U_{\pi}$ & $\mathds{1}$ & $U_{\pi}$ \\
\hline
\end{tabular}\label{table:corrections vs resource states}
\vspace*{-3mm}
\end{center}
\caption{Correction operations for even- and odd-parity resource states. Depending on which of the four outcomes (rows) is obtained, one of the four unitary corrections $\mathds{1}$, $U_{\pi}$, $U\subtiny{-1}{0}{\mathrm{P}}$, or $U\subtiny{-1}{0}{\mathrm{P}}U_{\pi}$ needs to be applied, depending on the resource state (columns) used.}
\end{table}
\noindent
such that
\begin{subequations}
\label{eq:Up action odd}
\begin{align}
    U\subtiny{-1}{0}{\mathrm{P}}\fket{1\subtiny{-1}{0}{B}}   &=
    \fdag{C}\f{B}\fket{1\subtiny{-1}{0}{B}}\,=\,\fket{1\subtiny{-1}{0}{C}},\\
    U\subtiny{-1}{0}{\mathrm{P}}\fket{1\subtiny{-1}{0}{\tilde{A}\pr}}   &=
    -\fdag{C}\fdag{B}\fket{1\subtiny{-1}{0}{\tilde{A}\pr}}\,=\,-\fket{1\subtiny{-1}{0}{B}}\fket{1\subtiny{-1}{0}{\tilde{A}\pr}}\fket{1\subtiny{-1}{0}{C}}.
\end{align}
\end{subequations}
One can confirm that the unitarity condition $U\subtiny{-1}{0}{\mathrm{P}}^{\dagger}U\subtiny{-1}{0}{\mathrm{P}}=U\subtiny{-1}{0}{\mathrm{P}}U\subtiny{-1}{0}{\mathrm{P}}^{\dagger}=\mathds{1}$ is satisfied using the anticommutation relations of~(\ref{eq:anticomm relations}). Bob may thus obtain the desired state
$\fket{\psi\suptiny{0}{0}{\mathrm{e}}}\subtiny{-1}{0}{B\nl \tilde{A}\pr}=\alpha\fket{0}+\beta\fket{1\subtiny{-1}{0}{B}}\fket{1\subtiny{-1}{0}{\tilde{A}\pr}}$ by applying the unitary $U\subtiny{-1}{0}{\mathrm{P}}U_{\pi}$ or just $U\subtiny{-1}{0}{\mathrm{P}}$, if the measurement outcome is $\fket{\Psi^{\pm}}\subtiny{-1}{0}{\tilde{A}\nl A}$ or $\fket{\Psi^{\mp}}\subtiny{-1}{0}{\tilde{A}\nl A}$, given that the resource state was $\fket{\Phi^{\pm}}\subtiny{-1}{0}{A\nl B}$.

When the state to be teleported has odd parity, $\fket{\psi}\subtiny{-1}{0}{\tilde{A}\nl \tilde{A}\pr}=\fket{\psi\suptiny{0}{0}{\mathrm{o}}}\subtiny{-1}{0}{\tilde{A}\nl \tilde{A}\pr}$, we have
\begin{align}
    \fket{\psi\suptiny{0}{0}{\mathrm{o}}}\subtiny{-1}{0}{\tilde{A}\nl \tilde{A}'}
    \fket{\Phi^{\pm}}\subtiny{-1}{0}{A\nl B}  &=\,
    \tfrac{1}{2}\Bigl[
        \fket{\Phi^{+}}\subtiny{-1}{0}{\tilde{A}\nl A}
        \bigl(\alpha\fket{1\subtiny{-1}{0}{\tilde{A}\pr}}\pm\beta\fket{1\subtiny{-1}{0}{B}}\bigr)
        \nonumber\\[1mm]
        &\ \ +
        \fket{\Phi^{-}}\subtiny{-1}{0}{\tilde{A}\nl A}
        \bigl(\alpha\fket{1\subtiny{-1}{0}{\tilde{A}\pr}}\mp\beta\fket{1\subtiny{-1}{0}{B}}\bigr)
        \nonumber\\[1mm]
        &\ \ \pm
        \fket{\Psi^{+}}\subtiny{-1}{0}{\tilde{A}\nl A}
        \bigl(\alpha\fket{1\subtiny{-1}{0}{B}}\fket{1\subtiny{-1}{0}{\tilde{A}\pr}}\pm\beta\fket{0}\bigr)
        \nonumber\\[1mm]
        &\ \ \pm
        \fket{\Psi^{-}}\subtiny{-1}{0}{\tilde{A}\nl A}
        \bigl(\alpha\fket{1\subtiny{-1}{0}{B}}\fket{1\subtiny{-1}{0}{\tilde{A}\pr}}\mp\beta\fket{0}\bigr)
        \Bigr].
        \label{eq:odd even}
\end{align}
For outcomes in the same parity sector as the resource state, the applied corrections are either trivial or correspond to $U_{\pi}$ from Eq.~(\ref{eq:Upi}). When the outcomes are in the odd-parity sector, we have to apply $U\subtiny{-1}{0}{\mathrm{P}}$ in addition, which acts as
\begin{subequations}
\label{eq:Up action even}
\begin{align}
    U\subtiny{-1}{0}{\mathrm{P}}\fket{0}   &=\,
    -\fdag{C}\fdag{B}\fket{0}\,=\,
    \fket{1\subtiny{-1}{0}{B}}\fket{1\subtiny{-1}{0}{C}},\\
    U\subtiny{-1}{0}{\mathrm{P}}\fket{1\subtiny{-1}{0}{B}}\fket{1\subtiny{-1}{0}{\tilde{A}\pr}}   &=\,
    \fdag{C}\f{B}\fket{1\subtiny{-1}{0}{B}}\fket{1\subtiny{-1}{0}{\tilde{A}\pr}}\,=\,-\fket{1\subtiny{-1}{0}{\tilde{A}\pr}}\fket{1\subtiny{-1}{0}{C}}.
\end{align}
\end{subequations}
Crucially, the combinations of outcomes and corrections, summarized in Table~\ref{table:corrections vs resource states}, are exactly the same as for the even-parity state $\fket{\psi\suptiny{0}{0}{\mathrm{e}}}\subtiny{-1}{0}{\tilde{A}\nl \tilde{A}'}$, such that Bob is not required to have information about the parity of the unknown state to successfully teleport it.


\subsubsection{Odd-parity resource states}

We can of course also consider the cases where the entangled resource state for the teleportation is an odd-parity state, $\fket{\phi}\subtiny{-1}{0}{A\nl B}=\fket{\Psi^{\pm}}\subtiny{-1}{0}{A\nl B}$. For a teleported state with even parity we then have
\begin{align}
    \fket{\psi\suptiny{0}{0}{\mathrm{e}}}\subtiny{-1}{0}{\tilde{A}\nl \tilde{A}'}
    \fket{\Psi^{\pm}}\subtiny{-1}{0}{A\nl B}  &=\,
    \tfrac{1}{2}\Bigl[
        \fket{\Phi^{+}}\subtiny{-1}{0}{\tilde{A}\nl A}
        \bigl(\alpha\fket{1\subtiny{-1}{0}{B}}\mp\beta\fket{1\subtiny{-1}{0}{\tilde{A}\pr}}\bigr)
        \nonumber\\[1mm]
        &\ \ +
        \fket{\Phi^{-}}\subtiny{-1}{0}{\tilde{A}\nl A}
        \bigl(\alpha\fket{1\subtiny{-1}{0}{B}}\pm\beta\fket{1\subtiny{-1}{0}{\tilde{A}\pr}}\bigr)
        \nonumber\\[1mm]
        &\ \ \pm
        \fket{\Psi^{+}}\subtiny{-1}{0}{\tilde{A}\nl A}
        \bigl(\alpha\fket{\!0\!}\mp\beta\fket{\!1\subtiny{-1}{0}{B}\!}\fket{\!1\subtiny{-1}{0}{\tilde{A}\pr}\!}\bigr)
        \nonumber\\[1mm]
        &\ \ \pm
        \fket{\Psi^{-}}\subtiny{-1}{0}{\tilde{A}\nl A}
        \bigl(\alpha\fket{\!0\!}\pm\beta\fket{\!1\subtiny{-1}{0}{B}\!}\fket{\!1\subtiny{-1}{0}{\tilde{A}\pr}\!}\bigr)
        \Bigr],
        \label{eq:even odd}
\end{align}
while an odd-parity state to be teleported results in
\begin{align}
    \fket{\psi\suptiny{0}{0}{\mathrm{o}}\!}\subtiny{-1}{0}{\tilde{A}\nl \tilde{A}'}
    \fket{\!\Psi^{\pm}\!}\subtiny{-1}{0}{A\nl B}  &=
    \tfrac{1}{2}\Bigl[
        -\fket{\Phi^{+}\!}\subtiny{-1}{0}{\tilde{A}\nl A}
        \bigl(\alpha\fket{1\subtiny{-1}{0}{B}\!}\fket{1\subtiny{-1}{0}{\tilde{A}\pr}\!}\mp\beta\fket{\!0\!}\bigr)
        \nonumber\\[1mm]
        &\ \ -
        \fket{\Phi^{-}}\subtiny{-1}{0}{\tilde{A}\nl A}
         \bigl(\alpha\fket{1\subtiny{-1}{0}{B}}\fket{1\subtiny{-1}{0}{\tilde{A}\pr}}\pm\beta\fket{0}\bigr)
        \nonumber\\[1mm]
        &\ \ \mp
        \fket{\Psi^{+}}\subtiny{-1}{0}{\tilde{A}\nl A}
        \bigl(\alpha\fket{1\subtiny{-1}{0}{\tilde{A}\pr}}\mp\beta\fket{1\subtiny{-1}{0}{B}}\bigr)
        \nonumber\\[1mm]
        &\ \ \mp
        \fket{\Psi^{-}}\subtiny{-1}{0}{\tilde{A}\nl A}
        \bigl(\alpha\fket{1\subtiny{-1}{0}{\tilde{A}\pr}}\pm\beta\fket{1\subtiny{-1}{0}{B}}\bigr)
        \Bigr].
        \label{eq:odd odd}
\end{align}
From Eqs.~(\ref{eq:Up action odd}) and~(\ref{eq:Up action even}), we see that the corresponding combinations of outcomes and corrections (for both $\fket{\psi\suptiny{0}{0}{\mathrm{e}}}\subtiny{-1}{0}{\tilde{A}\nl \tilde{A}'}$ and $\fket{\psi\suptiny{0}{0}{\mathrm{o}}}\subtiny{-1}{0}{\tilde{A}\nl \tilde{A}'}$) are the same regardless of the parity of the teleported state but of course depend on the specific resource state used, as summarized in Table~\ref{table:corrections vs resource states}.


\subsection{Implementation via fermionic Gaussian operations}\label{sec:impl ferm Gauss}

It is interesting to note that the whole teleportation protocol can be implemented via fermionic Gaussian operations. For the correction operation $U_{\pi}$ this is easy to see since its generator $\fdag{B}\f{B}$ is quadratic in the mode operators (of mode $B$). For the operator $U\subtiny{-1}{0}{\mathrm{P}}$, this is also the case, which can be seen in the following way. First, note that $U\subtiny{-1}{0}{\mathrm{P}}$ is Hermitian, $U\subtiny{-1}{0}{\mathrm{P}}^{\dagger}=U\subtiny{-1}{0}{\mathrm{P}}$. Since this implies that $U\subtiny{-1}{0}{\mathrm{P}}^{2}=\mathds{1}$, the unitary $U\subtiny{-1}{0}{\mathrm{P}}$ can be considered to coincide with the Hamiltonian generating the unitary up to a global phase. That is, we can define $H\subtiny{-1}{0}{\mathrm{P}}:=\tfrac{\pi}{2}(U\subtiny{-1}{0}{\mathrm{P}}-\mathds{1})$ and calculate
\begin{align}
    e^{-i\,H\subtiny{-1}{0}{\mathrm{P}}}  &=\,
    \sum\limits_{n=0}^{\infty}\frac{(-i\,H\subtiny{-1}{0}{\mathrm{P}})^{n}}{n!}\,=\,
    e^{i\,\tfrac{\pi}{2}}\,
    \sum\limits_{n=0}^{\infty}\frac{(-i\,\tfrac{\pi}{2}\,U\subtiny{-1}{0}{\mathrm{P}})^{n}}{n!}\nonumber\\[1mm]
    &=\,
    i\Bigl(
        \sum\limits_{n=0}^{\infty}\frac{(-i\,\tfrac{\pi}{2}\,U\subtiny{-1}{0}{\mathrm{P}})^{2n}}{(2n)!}
        +
        \sum\limits_{n=0}^{\infty}\frac{(-i\,\tfrac{\pi}{2}\,U\subtiny{-1}{0}{\mathrm{P}})^{2n+1}}{(2n+1)!}
    \Bigr)\nonumber\\[1mm]
    &=\,
    i\Bigl(
        \mathds{1}\sum\limits_{n=0}^{\infty}\frac{(-i\,\tfrac{\pi}{2})^{2n}}{(2n)!}
        +
        U\subtiny{-1}{0}{\mathrm{P}}
        \sum\limits_{n=0}^{\infty}\frac{(-i\,\tfrac{\pi}{2})^{2n+1}}{(2n+1)!}
    \Bigr)\nonumber\\[1mm]
    &=\,
    i\Bigl(
        \mathds{1}\cos\bigl(\tfrac{\pi}{2}\bigr)
        -i
        U\subtiny{-1}{0}{\mathrm{P}}
        \sin\bigl(\tfrac{\pi}{2}\bigr)
    \Bigr)\,=\,U\subtiny{-1}{0}{\mathrm{P}}\,.
    \label{eq:scenario 1 unitary generator}
\end{align}
Since both operators $U_{\pi}$ and $U\subtiny{-1}{0}{\mathrm{P}}$ are quadratic in the mode operators, each individually, and hence also their combination $U\subtiny{-1}{0}{\mathrm{P}}U_{\pi}$ are fermionic Gaussian operations. And since the Bell states are Gaussian~\cite{SpeeSchwaigerGiedkeKraus2018} the Bell measurement is a fermionic Gaussian operation~\cite{Bravyi2005}. We thus see that teleportation can be carried out by Gaussian means: sharing $1$ fbit, performing a Gaussian measurement, sending classical information (2 bits encoding the outcome of Alice's measurement) from Alice to Bob, and applying fermionic Gaussian corrections depending on the bit values.

Here, we note that the latter correction operations also require the availability of an auxiliary mode $C$. We have assumed this mode to be in the ground state initially, but no part of the teleportation protocol depends on the particular initial state and final state of this mode, or whether one even knows which state it is. The preparation of this mode hence does not require similar levels of control as the preparation of the entangled resource states. At the same time, this means that the auxiliary mode can be reused arbitrarily many times without resetting it to a particular state in between. Consequently, we do not consider the availability of this mode to be a resource requirement on the same footing as the other resources used for teleportation.


\vspace*{-1.5mm}
\subsection{Teleporting `one qubit of quantum information' -- the role of mode $\tilde{A}'$}\label{subsubsec:role of mode Apr}
\vspace*{-1.5mm}

Let us now more carefully discuss the purpose of the explicit inclusion of the mode $\tilde{A}\pr$ in our previous calculations. As we have already mentioned in Sec.~\ref{sec:fermions vs qubits}, using the teleportation protocol as outlined above just to transfer information about mode $\tilde{A}$ could be considered to be a waste of resources. Parity superselection constrains the state of mode $\tilde{A}$ to be of the form of Eq.~(\ref{eq:single mode state}), i.e., diagonal in the occupation number basis, and hence a classical state. However, there are two ways in which the protocol above can nonetheless be seen as transferring quantum information, both of which rely on the mode $\tilde{A}\pr$.

On the one hand, the fermionic single-mode teleportation protocol can be considered as \emph{entanglement swapping} from the modes $\tilde{A}$ and $\tilde{A}\pr$ to the modes $B$ and $\tilde{A}\pr$, regardless of who is controlling mode $\tilde{A}\pr$. If the modes $\tilde{A}$ and $\tilde{A}\pr$ are initially in an entangled state, then the modes $B$ and $\tilde{A}\pr$ are in that very same entangled state after the teleportation protocol. More generally, this is true for any arbitrary state $\rho\subtiny{0}{0}{\tilde{A}\nl \tilde{A}\pr}$ of these modes, since the details of the protocol (for fixed resource state) do not depend on the parity of the teleported state, and any state $\rho\subtiny{0}{0}{\tilde{A}\nl \tilde{A}\pr}$ must be a convex mixture of even- and odd-parity states of $\tilde{A}$ and $\tilde{A}\pr$. A fully classical information transfer whereby the mode $\tilde{A}$ is measured and the result is sent to Bob via a classical channel cannot achieve this, despite the fact that such a procedure would be able to transmit all locally available information about the mode $\tilde{A}$. The described fermionic entanglement swapping protocol can thus be considered to transfer the equivalent of one qubit of quantum information in the sense of being able to transfer one half of a mode pair in an arbitrary (unknown and potentially entangled) state. The resources for this transfer are $1$ fbit of shared fermionic entanglement and communicating $2$ bits of classical information. In this sense, even individually accessible fbits are more useful than the corresponding classical mixtures, despite the fact that any number of consecutive local measurements restricted to single copies of fbits cannot distinguish between the two.

\begin{figure}[t!]
\begin{center}
\includegraphics[width=0.32\textwidth,trim={0cm 0cm 0cm 0cm},clip]{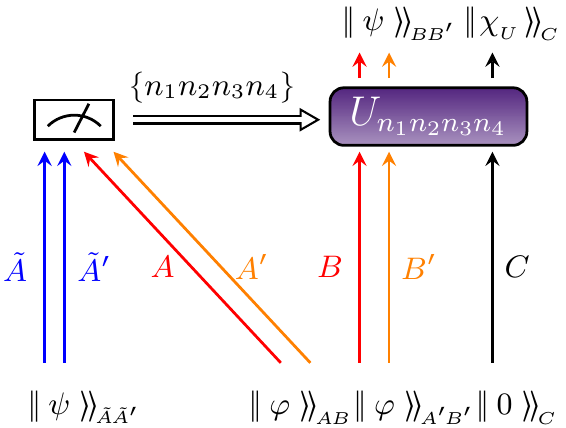}
\vspace*{-2.0mm}
\caption{Fermionic two-mode teleportation. In this scenario, teleporting quantum information encoded in the two-mode state $\fket{\!\psi\!}\subtiny{-1}{0}{\tilde{A}\nl \tilde{A}'}$, from modes $\tilde{A}$ and $\tilde{A}'$ to the modes $B$ and $B\pr$ requires two maximally entangled states $\fket{\phi}\subtiny{-1}{0}{A\nl B}$ and $\fket{\phi}\subtiny{-1}{0}{A\pr\nl B\pr}$ (2 fbits) and communicating four bits of classical information with values $n_1$, $n_2$, $n_3$, and $n_4$, where the bit pairs $\{n_{1},n_{2}\}$ and $\{n_{3},n_{4}\}$ encode the outcomes of the (independent) measurements on the mode pairs $\{\tilde{A},A\}$ and $\{\tilde{A}\pr,A\pr\}$, respectively. To complete the protocol a unitary operation $U_{n_{1}n_{2}n_{3}n_{4}}$ that depends on the bit values $n_{i}$ for $i=1,2,3,4$ is applied to the modes $B$ and $B\pr$, and to an auxiliary mode $C$. This may be realized as two consecutive operations $U_{n_{1}n_{2}}$ and $U_{n_{3}n_{4}}$ acting on the mode pairs $\{B,C\}$ and $\{B\pr,C\}$, respectively, and the state of mode $C$ does not need to be reset inbetween. The output state of the auxiliary mode $C$ is denoted as $\fket{\!\chi\subtiny{0}{0}{U}\!}\subtiny{-1}{0}{C}$ and depends on the local unitary operation $U_{n_{1}n_{2}n_{3}n_{4}}$ but remains separable from the other modes. The number of classical bits communicated from Alice to Bob can be reduced from $4$ to $2$, if non-Gaussian operations are used.}
\label{fig:twomodetele}
\end{center}
\end{figure}

On the other hand, one may argue that the equivalent of one qubit of quantum information should be defined in terms of the ability to encode the same complex amplitudes $\alpha$ and $\beta$ (with $|\alpha|^{2}+|\beta|^{2}=1$) as in a single-qubit state $\alpha\ket{0}+\beta\ket{1}$. Clearly, a single fermionic mode does not provide this ability, but two modes do. Therefore, one can realize the above single-mode protocol on both modes $\tilde{A}$ and $\tilde{A}\pr$ at once, in the way that `one qubit of quantum information' can be combined into full-fledged \emph{two-mode teleportation} by the iteration of the initially described entanglement swapping protocol.
That is, by using $2$ fbits entanglement, transferring $4$ bits of classical information and performing Gaussian operations (as discussed in Sec.~\ref{sec:impl ferm Gauss}), one may teleport the modes $\tilde{A}$ \emph{and} $\tilde{A}\pr$ as illustrated in Fig.~\ref{fig:twomodetele}. In this way, the complex amplitudes $\alpha$ and $\beta$ of any unknown two-mode state $\fket{\psi}\subtiny{-1}{0}{\tilde{A}\nl \tilde{A}'}$ (or, likewise, single-qubit state $\ket{\psi}\subtiny{-1}{0}{\tilde{A}\nl \tilde{A}'}$ in a dual-rail encoding) can be transferred.\\
\begin{table}[ht!]
\begin{center}
\renewcommand{\arraystretch}{1.55}
\begin{tabular}{ c|c|c|c| }
\cline{2-4}
& \multicolumn{1}{|c|}{\,One mode\,} & \multicolumn{2}{|c|}{Two modes}\\
\cline{1-4}
\multicolumn{1}{|c|}{fbits} & \multicolumn{1}{|c|}{1} & \multicolumn{1}{|c|}{\phantom{opera}1\phantom{opera}} & \multicolumn{1}{|c|}{\phantom{opera}2\phantom{opera}}\\
\cline{1-4}
\multicolumn{1}{|c|}{classical bits} & \multicolumn{1}{|c|}{2} & \multicolumn{1}{|c|}{2} & \multicolumn{1}{|c|}{2}\\
\cline{1-4}
\multicolumn{1}{|c|}{quantum channel} & \multicolumn{1}{|c|}{no} & \multicolumn{1}{|c|}{1 mode} & \multicolumn{1}{|c|}{no}\\
\hline
\end{tabular}\label{table:resource costs}
\vspace*{-3mm}
\end{center}
\caption{Resources and features of fermionic teleportation protocols for transmitting states of one or two modes. }
\end{table}

Indeed, one can even perform the two-mode teleportation protocol sharing only $2$ bits of classical information, if non-Gaussian operation are allowed, as Alice and Bob each locally perform a projective measurement of the parity of the resource state. For instance, if the resource state is $\fket{\Phi^{+}}\subtiny{-1}{0}{A\nl B}\fket{\Phi^{+}}\subtiny{-1}{0}{A\pr\nl B\pr}$, then an `even' outcome projects into $\tfrac{1}{\sqrt{2}}\bigl(\fket{0}+\fket{1\subtiny{-1}{0}{A},1\subtiny{-1}{0}{B},1\subtiny{-1}{0}{A\pr},1\subtiny{-1}{0}{B\pr}}\bigr)$, whereas an `odd' outcome results in $\tfrac{1}{\sqrt{2}}\bigl(\fket{1\subtiny{-1}{0}{A},1\subtiny{-1}{0}{B}}+\fket{1\subtiny{-1}{0}{A\pr},1\subtiny{-1}{0}{B\pr}}\bigr)$. In either case, Alice may then perform teleportation with a Bell measurement adapted to the measured parity and sending the usual $2$ classical bits (see, e.g., the example in Sec.~\ref{sec:two-mode tele NSSR} for comparison), while Bob learns the relevant parity from his local measurement. Thus it may seem as if an fbit is only half as powerful as an ebit, since two are needed to teleport a single qubit. However, this difference (almost) disappears if one allows us to teleport many fermionic modes at once. Then the even-parity sector of $n$ modes spans a $2^{n-1}$-dimensional Hilbert space uninhibited by parity superselection in which $n-1$ qubits can be encoded, and which can be faithfully teleported using $n$ fbits (whereas $n-1$ ebits would suffice without SSR). The resource costs of all three variants are summarized in Table~\ref{table:resource costs}.


\section{Fermionic teleportation subject to particle number superselection}\label{sec:particle number SSR teleportation}

\subsection{Non-fundamental superselection rules}

This far, we have viewed the task of fermionic teleportation as a fundamental problem, i.e., we have taken into account parity superselection but no other limitations. However, in practice, other restrictions such as non-fundamental SSRs typically do apply. In particular, we now want to discuss the influence of the particle number superselection rule (N-SSR).

Let us begin by noticing that it is less clear than with the P-SSR (see, e.g,. the discussion in~\cite{Friis2016a}), if the N-SSR is a fundamental restriction of Nature or not. On the one hand, we note that superpositions of different fermion numbers are not ruled out by charge conservation, much like superpositions of different energy eigenstates are not excluded by energy conservation. Instead, this can be viewed as an issue of not having available an appropriate reference frame; see, e.g., the discussion in Ref.~\cite[Sec.~IV]{Bartlett_review2007} or the argument by Aharonov and Susskind~\cite{AharonovSusskind1967}. At the same time, there does not appear to exist any process (to the best of our knowledge) that could result in a superposition of different electric charges. An example for a state with indefinite particle number sometimes referred to in this context is the BCS ground state~\cite{BardeenCooperSchrieffer1957}. However, the BCS ground state with indefinite electron number can be understood as convenient approximation of the actual physical state with fixed electron number~\cite{KrausWolfCiracGiedke2009}. Here, we therefore cannot conclusively answer the question if superpositions of different charges exist or not.

However, even if one were to adopt charge superselection axiomatically~\cite{StrocchiWightman1974}, one may of course consider species of uncharged fermions (both composite and fundamental). There, the question of the existence of pure states with indefinite particle number is tied to the question of the existence (or not) of Majorana fermions as fundamental objects in Nature. Although we cannot directly answer this question either, admittedly, the prospects of creating coherent superpositions of different numbers of fermions useful for quantum information processing are nevertheless daunting (to say the least) either way. For practical purposes, particle number superselection is hence a sensible restriction for practical implementations of fermionic teleportation such as in Ref.~\cite{OlofssonPottsBrunnerSamuelsson2020}.


\subsection{Single-mode teleportation \& particle number superselection}
\vspace*{-1.5mm}

To begin, it is interesting to put into perspective the usefulness of fbits as resource states for teleportation when constraints due to the N-SSR apply. In the context of the single-mode teleportation protocol discussed in Sec.~\ref{sec:scenario I}, particle number superselection implies that it is not possible to create or project into even-parity states of two fermionic modes $\tilde{A}$ and $A$ other than $\fket{0}$ and $\fket{1\subtiny{-1}{0}{\tilde{A}}}\fket{1\subtiny{-1}{0}{A}}$. More specifically, this means that the outset of the single-mode teleportation protocol is the restriction to the odd-parity state $\fket{\psi\suptiny{0}{0}{\mathrm{o}}\!}\subtiny{-1}{0}{\tilde{A}\nl \tilde{A}'}$ as the state to be teleported, and $\fket{\!\Psi^{\pm}\!}\subtiny{-1}{0}{A\nl B}$ as the shared resource state to achieve this. In addition, let us assume that the Bell measurement carried out by Alice can only result in states with definite particle number, i.e., the state of modes $\tilde{A}$ and $A$ will be projected into either $\fket{\Psi^{\pm}}\subtiny{-1}{0}{\tilde{A}\nl A}$, $\fket{0}\subtiny{-1}{0}{\tilde{A}\nl A}$, or $\fket{1\subtiny{-1}{0}{\tilde{A}}}\fket{1\subtiny{-1}{0}{A}}$. Consequently, it is instructive to write the initial joint state w.r.t. this choice of basis as
\begin{align}
    \fket{\psi\suptiny{0}{0}{\mathrm{o}}\!}\subtiny{-1}{0}{\tilde{A}\nl \tilde{A}'}
    \fket{\!\Psi^{\pm}\!}\subtiny{-1}{0}{A\nl B}  &=
    \tfrac{1}{2}\Bigl[
        -
        \fket{0}\subtiny{-1}{0}{\tilde{A}\nl A}\,
        \sqrt{2}\,\alpha\fket{1\subtiny{-1}{0}{B}\!}\fket{1\subtiny{-1}{0}{\tilde{A}\pr}\!}
        \nonumber\\[1mm]
        &\ \ \pm
        \fket{1\subtiny{-1}{0}{\tilde{A}}}\fket{1\subtiny{-1}{0}{A}}\,
         \sqrt{2}\,\beta\fket{0}\subtiny{-1}{0}{\tilde{A}\pr\nl B}
        \nonumber\\[1.5mm]
        &\ \ \mp
        \fket{\Psi^{+}}\subtiny{-1}{0}{\tilde{A}\nl A}
        \bigl(\alpha\fket{1\subtiny{-1}{0}{\tilde{A}\pr}}\mp\beta\fket{1\subtiny{-1}{0}{B}}\bigr)
        \nonumber\\[1mm]
        &\ \ \mp
        \fket{\Psi^{-}}\subtiny{-1}{0}{\tilde{A}\nl A}
        \bigl(\alpha\fket{1\subtiny{-1}{0}{\tilde{A}\pr}}\pm\beta\fket{1\subtiny{-1}{0}{B}}\bigr)
        \Bigr].
        \label{eq:odd odd particle number SSR}
\end{align}
As one can clearly see from this decomposition, single-mode teleportation can in this case only be successful if either $\fket{\Psi^{+}}\subtiny{-1}{0}{\tilde{A}\nl A}$ or $\fket{\Psi^{-}}\subtiny{-1}{0}{\tilde{A}\nl A}$ is obtained as the outcome on Alice's side, resulting in an average single-mode teleportation fidelity that is reduced by $50\%$ with respect to the case where no N-SSR applies. In principle, one may consider a more general scenario, where a measurement corresponding to a three-element POVM $\left\{P_{\Psi^+},P_{\Psi^-},P_{\mathrm{even\, parity}}\right\}$ is performed. In the case of the third outcome, the quantum information might still be present. However, it is delocalized between Alice and Bob, and we are not aware of any way to complete the transfer if particle number superselection applies.


\vspace*{-1.5mm}
\subsection{Two-mode teleportation \& particle number superselection}\label{sec:two-mode tele NSSR}
\vspace*{-1.5mm}

Let us now consider two-mode teleportation in the presence of particle superselection. The state to be teleported in this scenario is a two-mode state containing a single fermion, which can be considered as dual-rail encoding of a qubit. If we combine two fbits in the odd-parity sector (two single-fermion states as in the implementations proposed in Ref.~\cite{OlofssonPottsBrunnerSamuelsson2020} and as discussed in Sec.~\ref{subsubsec:role of mode Apr}) as resource states and naively perform the teleportation for each mode separately as before, then we see that the teleportation fidelity is further reduced to $25\%$, since the teleportation of either mode is only successful half the time (on average). However, as we shall see shortly, particle superselection does not intrinsically limit the fidelity in this way. Using the same resource state (a pair of two single-fermion fbits), the fidelity can be increased to $50\%$, and for other resource states (subject to particle superselection) one may even achieve $100\%$ teleportation fidelity.

To see, this, let us consider a different resource state for the modes $A$, $B$, $A\pr$ and $B\pr$ in a setup subject to particle superselection. Take, for instance, the state
\begin{align}
    \fket{\Psi^{+}_{\mathrm{R}}}\subtiny{-1}{0}{A\nl B\nl A\pr\nl B\pr} &=\,
    \tfrac{1}{\sqrt{2}}\bigl(
    \fket{1\subtiny{-1}{0}{A}}\fket{1\subtiny{-1}{0}{B\pr}}
    +\fket{1\subtiny{-1}{0}{B}}\fket{1\subtiny{-1}{0}{A\pr}}
    \bigr),
    \label{eq:four mode resource state}
\end{align}
and let Alice carry out a projective measurement on the modes $\tilde{A}$, $\tilde{A}\pr$, $A$, and $A\pr$ in the `basis' given by the four states
\begin{subequations}
\begin{align}
    \fket{\Phi_{\mathrm{R}}^{\pm}}\subtiny{-1}{0}{\tilde{A}\nl \tilde{A}\pr\nl A\nl A\pr} &=\,\tfrac{1}{\sqrt{2}}\bigl(
    \fket{1\subtiny{-1}{0}{\tilde{A}}}\fket{1\subtiny{-1}{0}{A}}
    \pm\fket{1\subtiny{-1}{0}{\tilde{A}\pr}}\fket{1\subtiny{-1}{0}{A\pr}}
    \bigr),\\[1mm]
    \fket{\Psi_{\mathrm{R}}^{\pm}}\subtiny{-1}{0}{\tilde{A}\nl \tilde{A}\pr\nl A\nl A\pr} &=\,\tfrac{1}{\sqrt{2}}\bigl(
    \fket{1\subtiny{-1}{0}{\tilde{A}}}\fket{1\subtiny{-1}{0}{A\pr}}
    \pm\fket{1\subtiny{-1}{0}{\tilde{A}\pr}}\fket{1\subtiny{-1}{0}{A}}
    \bigr).
\end{align}
\end{subequations}
Here, we have put `basis' in quotation marks, since these four states form a basis only of that subspace of the $2$-fermion subspace\footnote{The $2$-particle sector of the Fock space of $4$ fermionic modes is $6$-dimensional, but for two of these states, $\fket{1\subtiny{-1}{0}{\tilde{A}}}\fket{1\subtiny{-1}{0}{\tilde{A}\pr}}$ and $\fket{1\subtiny{-1}{0}{A}}\fket{1\subtiny{-1}{0}{A\pr}}$, the particle content of the subspace of modes $\tilde{A}$ and $\tilde{A}\pr$ is different from $1$.} of the four modes in question where there is exactly $1$ fermion in the modes $\tilde{A}$ and $\tilde{A}\pr$. Consequently, when a single-fermion state $\fket{\psi}\subtiny{-1}{0}{\tilde{A}\nl \tilde{A}'}=\fket{\psi\suptiny{0}{0}{\mathrm{o}}}\subtiny{-1}{0}{\tilde{A}\nl \tilde{A}'}$ is prepared for the modes $\tilde{A}$ and $\tilde{A}\pr$, we can write
\begin{align}
    \fket{\!\psi\!}\subtiny{-1}{0}{\tilde{A}\nl \tilde{A}'}\fket{\!\Psi^{+}_{\mathrm{R}}\!}\subtiny{-1}{0}{A\nl B\nl A\pr\nl B\pr}
    &=\tfrac{1}{2}\Bigl[
    \fket{\Psi_{\mathrm{R}}^{+}}\subtiny{-1}{0}{\tilde{A}\nl \tilde{A}'\nl A\nl A\pr}
        \bigl(\alpha\fket{\!1\subtiny{-1}{0}{B\pr}\!}-\beta\fket{\!1\subtiny{-1}{0}{B}\!}\bigr)\nonumber\\[1mm]
    &\ \ -
    \fket{\Psi_{\mathrm{R}}^{-}}\subtiny{-1}{0}{\tilde{A}\nl \tilde{A}'\nl A\nl A\pr}
        \bigl(\alpha\fket{\!1\subtiny{-1}{0}{B\pr}\!}+\beta\fket{\!1\subtiny{-1}{0}{B}\!}\bigr)\nonumber\\[1mm]
    &\ \ -
    \fket{\Phi_{\mathrm{R}}^{+}}\subtiny{-1}{0}{\tilde{A}\nl \tilde{A}'\nl A\nl A\pr}
        \bigl(\alpha\fket{\!1\subtiny{-1}{0}{B}\!}-\beta\fket{\!1\subtiny{-1}{0}{B\pr}\!}\bigr)\nonumber\\[1mm]
    &\ \ +
    \fket{\Phi_{\mathrm{R}}^{-}}\subtiny{-1}{0}{\tilde{A}\nl \tilde{A}'\nl A\nl A\pr}
        \bigl(\alpha\fket{\!1\subtiny{-1}{0}{B}\!}+\beta\fket{\!1\subtiny{-1}{0}{B\pr}\!}\bigr)\bigr].\nonumber
\end{align}
We thus see that the encoding of the teleported state, the preparation of the resource state, the Bell measurements, as well as any correction operations required on the modes $B$ and $B\pr$ can all in principle be carried out while respecting particle number (and charge) superselection, both globally and locally (with respect to the partition $\tilde{A}\nl \tilde{A}\pr|A\nl A\pr|B\nl B\pr$), achieving a teleportation fidelity of $100\%$.

However, we observe that the resource (state) for this teleportation is not a pair of fbits anymore. This can be understood in a simple way: Although both resource states, $\fket{\Psi^{+}_{\mathrm{R}}}\subtiny{-1}{0}{A\nl B\nl A\pr\nl B\pr}$ and $\fket{\Psi^{+}}\subtiny{-1}{0}{A\nl B}\fket{\Psi^{+}}\subtiny{-1}{0}{A\pr\nl B\pr}$, are pure states with the same particle content ($2$ fermions), and can hence be transformed into each other by global (on $A$, $B$, $A\pr$, $B\pr$) particle-number conserving unitaries, this cannot be achieved by unitaries acting \emph{locally} with respect to the bipartition $AA\pr|B\nl B\pr$. To see this, simply note that the reduced states of the modes $A$ and $A\pr$ have different ranks for the different resource states. That is, both states are entangled w.r.t. this cut, but (\emph{it seems}) not equally strongly (w.r.t. to an entanglement measure suitable to the applicable SSR). Nevertheless, if two fbits  $\fket{\Psi^{+}}\subtiny{-1}{0}{A\nl B}\fket{\Psi^{+}}\subtiny{-1}{0}{A\pr\nl B\pr}$ are used as a resource, one may still achieve $50\%$ fidelity. If Alice performs a projective measurement of the total particle number in modes $A$ and $A\pr$ before performing the Bell measurement, this will result in the state $\fket{\Psi^{+}_{\mathrm{R}}}\subtiny{-1}{0}{A\nl B\nl A\pr\nl B\pr}$ in half the cases (when there is $1$ particle in the modes $A$ and $A\pr$), and in separable states $\fket{\!1\subtiny{-1}{0}{B},1\subtiny{-1}{0}{B\pr}\!}$ (when there are no particles in the modes $A$ and $A\pr$) and $\fket{\!1\subtiny{-1}{0}{A},1\subtiny{-1}{0}{A\pr}\!}$ (two particles in the modes $A$ and $A\pr$) otherwise.

In other words, problems arise from using resource states whose marginals have support in different superselection sectors. All restrictions disappear, of course, if all logical qubits are locally supported in a subspace of fixed particle number. Then N-SSR does not restrict any logical operations, the projection on each of the four logical Bell states is permitted, and standard teleportation (of logical qubits) works as usual. The limitation of the fidelity due to particle number superselection in potential experimental settings as discussed in detail in Ref.~\cite{OlofssonPottsBrunnerSamuelsson2020} is hence more a practical (but nonetheless very challenging) problem of determining ways to prepare states like $\fket{\Psi^{+}_{\mathrm{R}}}\subtiny{-1}{0}{A\nl B\nl A\pr\nl B\pr}$ from Eq.~(\ref{eq:four mode resource state}) directly (rather than by post-selection after preparing two fbits). In particular, the preparation of states like $\fket{\Psi^{+}_{\mathrm{R}}}\subtiny{-1}{0}{A\nl B\nl A\pr\nl B\pr}$ for spin systems is challenging when the interaction between individual spins is weak. Nevertheless, this limitation can be overcome, for instance, in experiments based on pseudo-spins in double-well quantum dots, e.g., as in Refs.~\cite{kim2011ultrafast, delteil2016generation, nichol2017}.


\section{Implications for fermionic entanglement}\label{sec:implications}

In this section, we want to relate our previous observations about fermionic teleportation with the quantification of entanglement subject to SSRs. In particular, we aim here to contrast the notion of superselected entanglement of formation (EOF, as discussed in Sec.~\ref{sec:entanglement of ferm modes}) with a state's usefulness for teleportation.

For pure states, i.e., $1$ or $2$ fbits, this appears to be rather straightforward. The superselected EOF of $n$ (pure) fbits is equal to $n$, and we can refer to Table~\ref{table:resource costs} for the corresponding resources for different tasks. However, the superselected EOF allows for the notion of `mixed maximally entangled' (MME) fermionic states~\cite{DArianoManessiPerinottiTosini2014a}. Take, for instance, the MME state
\begin{align}
    \rho\subtiny{-1}{0}{A\nl B}\suptiny{0}{0}{\mathrm{MME}}  &=\,
    \tfrac{1}{2}\fket{\Phi^{+}}\!\!\fbra{\Phi^{+}}+\tfrac{1}{2}\fket{\Psi^{+}}\!\!\fbra{\Psi^{+}}.
    \label{eq:MME state two modes}
\end{align}
Because the two parity subspaces do not mix, one fbit is required per copy to create $\rho\subtiny{-1}{0}{A\nl B}\suptiny{0}{0}{\mathrm{MME}}$ and the (parity) superselected EOF evaluates\footnote{Here, we choose the logarithm to base $2$ in the von~Neumann entropy.} to $\mathcal{E}_{\mathrm{oF}}(\rho\subtiny{-1}{0}{A\nl B}\suptiny{0}{0}{\mathrm{MME}})=\log(2)=1$. The entanglement of $\rho\subtiny{-1}{0}{A\nl B}\suptiny{0}{0}{\mathrm{MME}}$ can thus be considered to be maximal in this sense. \emph{But are MME states useful for teleportation?} In the following, we will discuss this question in more detail for the P-SSR and the N-SSR.


\subsection{Teleportation using mixed maximally entangled states for P-SSR}\label{sec:Teleportation using mixed maximally entangled states for P-SSR}

Let us consider fermionic teleportation using the state $\rho\subtiny{-1}{0}{A\nl B}\suptiny{0}{0}{\mathrm{MME}}$ as a resource state for teleporting the state of mode $\tilde{A}$ as illustrated in Fig.~\ref{fig:telescenario1}. Note that the mode $\tilde{A}$ may be entangled with another single mode $\tilde{A}\pr$ (or even with multiple other modes). If the state of modes $\tilde{A}$ and $\tilde{A}\pr$ has even parity and is given by $\fket{\psi\suptiny{0}{0}{\mathrm{e}}}\subtiny{-1}{0}{\tilde{A}\nl \tilde{A}\pr}$, and the outcome of the Bell-measurement in the basis $\{\fket{\Phi^{\pm}}\subtiny{-1}{0}{\tilde{A}\nl A},\fket{\Psi^{\pm}}\subtiny{-1}{0}{\tilde{A}\nl A}\}$ gives the outcome $\fket{\Phi^{+}}\subtiny{-1}{0}{\tilde{A}\nl A}$, Eqs.~(\ref{eq:even even}) and~(\ref{eq:even odd}) allow us to conclude that the state of modes $B$ and $\tilde{A}\pr$ prior to any corrections is an equally weighted mixture of $\alpha\fket{0}+\beta\fket{1\subtiny{-1}{0}{B}}\fket{1\subtiny{-1}{0}{\tilde{A}\pr}}$ and $\alpha\fket{1\subtiny{-1}{0}{B}}-\beta\fket{1\subtiny{-1}{0}{\tilde{A}\pr}}$. In particular, this means that the reduced state of mode $B$ is given by $\tfrac{1}{2}\fket{0}\!\!\fbra{0}+\tfrac{1}{2}\fket{1\subtiny{-1}{0}{B}}\!\!\fbra{1\subtiny{-1}{0}{B}}$ and is hence maximally mixed. This means, no information whatsoever about the teleported state is locally available in mode $B$. However, if we consider the joint state of modes $B$ and $\tilde{A}\pr$, we see that \emph{all} information about the teleported state is still available. That is, a joint parity measurement on both modes projects either into the state $\alpha\fket{0}+\beta\fket{1\subtiny{-1}{0}{B}}\fket{1\subtiny{-1}{0}{\tilde{A}\pr}}$, if the result was `even', or into the state $\alpha\fket{1\subtiny{-1}{0}{B}}-\beta\fket{1\subtiny{-1}{0}{\tilde{A}\pr}}$, if the result was `odd'. In the former case, one has already retrieved the desired state. In the latter case, one applies the unitary correction $U\subtiny{-1}{0}{\mathrm{P}}$ to complete the teleportation. However, if the mode $\tilde{A}\pr$ is under Alice's control, then the required global parity measurement is itself a non-local operation which can create entanglement and thus the MME state is not necessary for teleportation in this scenario. But if the mode to be teleported is known to be only entangled with modes under Bob's control, then the parity measurement can be implemented locally and the teleportation can be completed using the shared MME state. We refer to this process as `\emph{subsystem swap}' in the following. As this example illustrates (and as can easily be confirmed for other combinations of resource states and teleported states), \emph{MME states can indeed be useful for `teleportation'}, if only for very specific tasks and keeping in mind important caveats that we discuss in the following.

The first difference to using pure maximally entangled states manifests in the amount of information that is available locally about the teleported state. That is, the teleportation protocol using the two-mode MME state (2MMES) becomes useful only if the joint parity of the modes $\tilde{A}\tilde{A}\pr$ is known, and one has access also to the second mode $\tilde{A}\pr$ to perform a joint (and non-destructive as well as non-particle number resolving) parity measurement on the modes $\tilde{A}\pr$ and $B$.

Besides the subsystem swap, there is a second, more standard teleportation related task that the 2MMES allows us to perform, proving its value as a \emph{non-local} resource. That is, it allows us to locally convert an fbit into an ebit. This state transformation from, say, $\rho\subtiny{-1}{0}{A\nl B}\suptiny{0}{0}{\mathrm{MME}}$ and $\fket{\Psi^{+}}\subtiny{-1}{0}{A\pr B\pr}$ to $\left(\fket{0101}\subtiny{-1}{0}{AA\pr BB\pr}+\fket{1010}\subtiny{-1}{0}{AA\pr BB\pr}\right)/\sqrt{2}$, is realized if both Alice and Bob measure their respective local parity and obtain an even result. (For the other outcomes, the resulting final state is also an ebit.) That such a transformation is not possible by local parity-constrained operations is confirmed by their respective Schmidt coefficients. As analyzed by~\cite{SchuchVerstraeteCirac2004}, for states subject to local and global SSRs, the vector of Schmidt coefficients governing local state transformations for bipartite pure states of distinguishable quantum systems~\cite{Nielsen1999} must be replaced by a \emph{set} of several vectors, one for each SSR-sector at $A$, and state transformations are only possible if \emph{all} (in our case: both) Schmidt vectors of the target state majorize those of the source state. But for the case at hand the source state has two one-dimensional vectors $\{(1/2), (1/2)\}$, while the target state has one two-dimensional Schmidt vector $\{(1/2,1/2),(\varnothing)\}$ and thus the transformation is impossible with local means (respecting the SSR). Note that the 2MMES allows us to implement both tasks (single-mode teleportation and conversion of an fbit into an ebit) exactly and with probability 1 (as before, provided that one is able to perform non-Gaussian operations.)

The remarkable aspect here is not that mixed states \emph{can} be used for (special-purpose) teleportation but that a state which could \textemdash\ in the absence of SSRs \textemdash\ be generated by local operations and classical communication (LOCC) allows us to realize non-trivial entanglement transformations. Note, however, that both tasks are feasible locally in the qubit setting, but both require operations forbidden by the SSR to achieve them locally and the 2MMES allows us to lift these parity-imposed restrictions without violating the SSR.

Another way to interpret the second task is to note that the four-mode state $\rho\subtiny{-1}{0}{A\nl B}\suptiny{0}{0}{\mathrm{MME}}\wedge\fket{\varphi}\!\!\fbra{\varphi}\subtiny{0}{0}{A\pr\nl B\pr}$, where $\fket{\varphi}\subtiny{0}{0}{A\pr\nl B\pr}$ is a pure fbit, can be converted to an ebit encoded in four fermionic modes by LOCC. (Note that this is \emph{not} reversible, as by the Schmidt-vector argument used before the fbit and the fermionic ebit are incomparable.) Moreover, we observe that the latter resource state $\rho\subtiny{-1}{0}{A\nl B}\suptiny{0}{0}{\mathrm{MME}}\wedge\fket{\varphi}\!\!\fbra{\varphi}\subtiny{0}{0}{A\pr\nl B\pr}$ is itself an \emph{MME state of $4$ modes}, i.e., a mixture of $2$ pure two-fbit states in the two different parity sectors.

The second difference between pure and mixed maximally entangled fermionic states lies in the security of the teleportation. That is, two (pure) fbits allow for violating a Bell inequality~\cite{DasenbrookBowlesBohrBraskHoferFlindtBrunner2016}, and hence for authentication, whereas any number of copies of $2$-mode MME states as in Eq.~(\ref{eq:MME state two modes}) alone does not. To see this more clearly, note that the two-qubit equivalent $\tilde{\rho}\subtiny{-1}{0}{A\nl B}$ (not subject to any SSRs) of $\rho\subtiny{-1}{0}{A\nl B}\suptiny{0}{0}{\mathrm{MME}}$ is separable (which can easily be checked via the Peres-Horodecki criterion~\cite{Peres1996, HorodeckiMPR1996}), and therefore so are two copies of $\tilde{\rho}\subtiny{-1}{0}{A\nl B}$. Therefore, no Bell inequality can be violated by $\tilde{\rho}\subtiny{-1}{0}{A\nl B}$, or by any number of copies of $\tilde{\rho}\subtiny{-1}{0}{A\nl B}$. This is so because SSRs further restrict the measurable operators that may appear in a Bell inequality. Consequently, the superselected state $\rho\subtiny{-1}{0}{A\nl B}\suptiny{0}{0}{\mathrm{MME}}$ (or two copies of it) can also not violate a Bell inequality.

Nevertheless, authentication is possible (albeit, at a higher price) if one uses the $4$-mode MME state for teleportation, one simply has to sacrifice twice as many (as compared to the situation using $2$ fbits per teleported qubit) of the resource states for authentication to retrieve the same number of fbit pairs. Just recall that, also with pure states one has to collect statistics on measurements of sufficiently many entangled resource states to violate a Bell inequality. The choice between pure and mixed maximally entangled states hence comes down to a matter of efficiency of the authentication.

A comparison of the usefulness of MME states and fbits is shown in Table~\ref{table:PSSR teleportation summary}. In summary, we can say that for some very specific tasks, MME states seem to have some usefulness comparable with fbits, and this is reflected in the matching values of EOF. However, in general they are clearly less useful. In particular, the difference in the potential to violate Bell inequalities is not captured by the (superselected) EOF. Nevertheless, MME states allow us to perform some tasks made locally impossible by superselection. Finally, let us briefly discuss the extension of the ideas of MME states and MME-based teleportation to other SSRs.

\begin{table}[ht!]
\begin{center}
\renewcommand{\arraystretch}{1.55}
\begin{tabular}{ c|c|c|c|c|c| }
\cline{2-6}
& \multicolumn{1}{|c|}{\,$\substack{\text{$2$-mode}\\[0.5mm] \text{MME}}$\,} & \multicolumn{1}{|c|}{\,1 fbit\,} & \multicolumn{1}{|c|}{\,$\substack{\text{$4$-mode}\\[0.5mm] \text{ferm. ebit}}$\,} & \multicolumn{1}{|c|}{\,$\substack{\text{$4$-mode}\\[0.5mm] \text{MME}}$\,} & \multicolumn{1}{|c|}{\,2 fbits\,}\\
\hline
\multicolumn{1}{|c|}{\,EOF\,} & \multicolumn{1}{|c|}{\,1\,} & \multicolumn{1}{|c|}{\,1\,} & \multicolumn{1}{|c|}{\,2\,} & \multicolumn{1}{|c|}{\,2\,} & \multicolumn{1}{|c|}{\,2\,}\\
\hline
\multicolumn{1}{|c|}{\,$\substack{\text{subsystem}\\[0.5mm] \text{swap}}$\,} & \multicolumn{1}{|c|}{\,1 mode$^{\dagger}$\,} & \multicolumn{1}{|c|}{\,1 mode\,} & \multicolumn{1}{|c|}{\,---\,}  & \multicolumn{1}{|c|}{\,2 modes$^{\dagger}$\,} & \multicolumn{1}{|c|}{\,2 modes\,}\\
\hline
\multicolumn{1}{|c|}{\,$\substack{\text{teleportation}\\[0.5mm] \# \,\text{of qubits}}$\,} & \multicolumn{1}{|c|}{\,0\,} & \multicolumn{1}{|c|}{\,0\,} & \multicolumn{1}{|c|}{\,1\,} & \multicolumn{1}{|c|}{\,1$^{\dagger}$\,} & \multicolumn{1}{|c|}{\,1\,}\\
\hline
\multicolumn{1}{|c|}{\,$\substack{\text{Bell inequ.}\\[0.5mm] \text{violation}}$\,} & \multicolumn{1}{|c|}{\,No\,} &\multicolumn{1}{|c|}{\,Yes$^{*}$\,} &  \multicolumn{1}{|c|}{\,Yes\,} & \multicolumn{1}{|c|}{\,Yes$^{*\dagger}$\,} & \multicolumn{1}{|c|}{\,Yes\,}\\
\hline
\end{tabular}\label{table:PSSR teleportation summary}
\vspace*{-4mm}
\end{center}
\caption{Comparison of ebits (maximally entangled state of four fermionic modes with fixed local parity), fbits, and MME states in terms of entanglement of formation [EOF (in units of fbits)], subsystem swapping capacity in terms of the number of modes whose state can be swapped (as described in the text, cf. also Fig.~\ref{fig:telescenario1}, but with mode $\tilde{A}\pr$ already in Bob's possession), number of qubits (two-dimensional subspaces) that can be teleported, and potential for Bell inequality violation (given sufficiently many copies). Superscripts denote that the required operations are non-Gaussian ($\dagger$) or have to be performed coherently on two copies of the state ($^{*}$), requiring a quantum memory. The states are strictly more entangled from left to right, since fermionic LOCC allow us to go to the left neighbour but not to the right -- except that the fermionic ebit is incomparable with the fbit and 2MMES (no conversion possible either way).
}
\end{table}


\subsection{Teleportation using mixed maximally entangled states for N-SSR}\label{sec:Teleportation using mixed maximally entangled states for N-SSR}

An obvious question that arises then concerns the usefulness (and existence) of MME states for other SSRs, in particular, for particle numbers superselection. The four-mode MME states encountered in the previous section allow for a $100\%$ teleportation fidelity when only parity superselection applies. The crucial element is a final projective measurement of the system's parity. However, when the N-SSR is in place, the state $\rho\subtiny{-1}{0}{A\nl B}\suptiny{0}{0}{\mathrm{MME}}$ is no longer allowed by the SSR. Replacing the even Bell state by the statistical mixture of its two components $\fket{0}$ and $\fket{1\subtiny{-1}{0}{A}}\fket{1\subtiny{-1}{0}{B}}$ turns $\rho\subtiny{-1}{0}{A\nl B}\suptiny{0}{0}{\mathrm{MME}}$ into a state that is neither maximally entangled, nor useful for teleportation. Nevertheless, this does not mean one cannot consider other states that correspond to MME states in the presence of the N-SSR (or, indeed, any SSR).

Let us consider a scenario where particle number superselection applies and we wish to teleport the state of $1$ qutrit encoded in three fermionic modes labelled $\tilde{A}$, $\tilde{A}\pr$ and $\tilde{A}\prpr$. This can be done by encoding the qutrit in the single-particle sector, spanned by the vectors $\fket{1\subtiny{-1}{0}{\tilde{A}}}$, $\fket{1\subtiny{-1}{0}{\tilde{A}\pr}}$, and $\fket{1\subtiny{-1}{0}{\tilde{A}\prpr}}$, or in the two-particle sector, spanned by the vectors $\fket{1\subtiny{-1}{0}{\tilde{A}}1\subtiny{-1}{0}{\tilde{A}\pr}}$, $\fket{1\subtiny{-1}{0}{\tilde{A}\pr}1\subtiny{-1}{0}{\tilde{A}\prpr}}$, and $\fket{1\subtiny{-1}{0}{\tilde{A}\pr}1\subtiny{-1}{0}{\tilde{A}\prpr}}$. As a resource for teleportation we can then use any $6$-mode state (say, of modes $A$, $B$, $A\pr$, $B\pr$, $A\prpr$, and $B\prpr$) whose particle number is fixed both globally (to $2$, $3$, or $4$ particles) and locally (to either $1$ or $2$ particles).

For instance, let us adopt the notation $\fket{n;j,k}$ for a state of $n$ particles of which $j$ particles are in the subspace of $A$-modes $A$, $A\pr$, $A\prpr$, and $k$ particles in the subspace of the $B$-modes $B$, $B\pr$, $B\prpr$. Then, for example, one of the following states can be used for teleportation:
\begin{subequations}
\begin{align}
    \fket{2;1,1} & =\tfrac{1}{\sqrt{3}}\bigl(
        \fket{\!1\subtiny{-1}{0}{A},1\subtiny{-1}{0}{B}\!}
        +\fket{\!1\subtiny{-1}{0}{A\pr},1\subtiny{-1}{0}{B\pr}\!}
        +\fket{\!1\subtiny{-1}{0}{A\prpr},1\subtiny{-1}{0}{B\prpr}\!}\bigr),\\[1mm]
    \fket{3;1,2} & =\tfrac{1}{\sqrt{3}}\bigl(
        \fket{\!1\subtiny{-1}{0}{A},1\subtiny{-1}{0}{B},1\subtiny{-1}{0}{B\pr}\!}
        +\fket{\!1\subtiny{-1}{0}{A\pr},1\subtiny{-1}{0}{B},1\subtiny{-1}{0}{B\prpr}\!}\nonumber\\
        &\ \ \ \ \ \ +\fket{\!1\subtiny{-1}{0}{A\prpr},1\subtiny{-1}{0}{B\pr},1\subtiny{-1}{0}{B\prpr}\!}\bigr),\\[1mm]
    \fket{3;2,1} & =\tfrac{1}{\sqrt{3}}\bigl(
        \fket{\!1\subtiny{-1}{0}{A},1\subtiny{-1}{0}{A\pr},1\subtiny{-1}{0}{B}\!}
        +\fket{\!1\subtiny{-1}{0}{A},1\subtiny{-1}{0}{A\prpr},1\subtiny{-1}{0}{B\pr}\!}\nonumber\\
        &\ \ \ \ \ \ +\fket{\!1\subtiny{-1}{0}{A\pr},1\subtiny{-1}{0}{A\prpr},1\subtiny{-1}{0}{B\prpr}\!}\bigr),\\[1mm]
    \fket{4;2,2} & =\tfrac{1}{\sqrt{3}}\bigl(
        \fket{\!1\subtiny{-1}{0}{A},1\subtiny{-1}{0}{A\pr},1\subtiny{-1}{0}{B},1\subtiny{-1}{0}{B\pr}\!}
        +\fket{\!1\subtiny{-1}{0}{A},1\subtiny{-1}{0}{A\prpr},1\subtiny{-1}{0}{B},1\subtiny{-1}{0}{B\prpr}\!}\nonumber\\
        &\ \ \ \ \ \ +\fket{\!1\subtiny{-1}{0}{A\pr},1\subtiny{-1}{0}{A\prpr},1\subtiny{-1}{0}{B\pr},1\subtiny{-1}{0}{B\prpr}\!}\bigr),
\end{align}
\end{subequations}
While any of these states can be used to teleport $1$ qutrit, we can also consider an arbitrary incoherent mixture of any of these four states as a mixed entangled resource state for teleportation. Such a teleportation protocol works in the following way: Alice and Bob share the mixed entangled resource state, Alice receives the $A$-modes, and Bob receives the $B$-modes. Alice then performs a projective measurement of the particle number on the $A$-modes before performing an appropriate Bell measurement on the $3\times3$-dimensional subspace of the $\tilde{A}$ and $A$-modes corresponding to the particle number of her encoded state and the result of the initial projective measurement on the $A$-modes. The result of the Bell measurement is communicated to Bob, who makes a similar projective measurement of the particle number on the $B$-modes and applies a correction depending on the classically communicated outcome of the Bell measurement.

As before for the parity SSR, the remarkable aspect lies not in the fact that mixed states can be used for teleportation in this way, but in the fact that there is no pure state of $6$ modes subject to particle number superselection that could do better than teleporting a single qutrit (or $\log_{2}3$ qubits) with unit fidelity (whereas a $6$-qubit state could be used to teleport $3$ qubits with the same fidelity). The maximum of $\log_{2}3$ qubits is simply the maximum dimension $d\subtiny{0}{0}{\mathrm{SSR}}\suptiny{0}{0}{\mathrm{max}}$ of any subspace of $3$ modes with fixed particle number. In general, the subspace dimension corresponding to $k$ particles in $n$ modes is $\tbinom{n}{k}$ and hence
\begin{align}
    d\subtiny{0}{0}{\mathrm{SSR}}\suptiny{0}{0}{\mathrm{max}}   &=\,
    \begin{cases}
        \binom{n}{n/2} & \text{if}\ $n$\ \text{even}\\[1mm]
        \binom{n}{(n-1)/2} & \text{if}\ $n$\ \text{odd}
    \end{cases}.
\end{align}
We thus see also for N-SSR that there exist MME states, and that these can also be useful for teleportation in terms of the number of transferred qubits, albeit with a reduced ability to violate Bell inequalities, as discussed in Sec.~\ref{sec:Teleportation using mixed maximally entangled states for P-SSR}. Moreover, analogous arguments can be made for any SSR. It is further interesting to remark that for $n$ even and large, $\log_2(d\subtiny{0}{0}{\mathrm{SSR}}\suptiny{0}{0}{\mathrm{max}})$ approaches $n$ (by Stirling's formula), i.e., $n$ modes allow for $n$ not-SSR-inhibited qubits asymptotically [up to $\log (n)$ corrections].


\section{Discussion}\label{sec:discussion}

We have reviewed quantum teleportation in a setting where quantum information is encoded in the modes of a fermionic quantum field. As we have discussed, differences to standard qubit-based teleportation arise due to parity superselection, which influences both the encoding of quantum information in the state space, as well as the allowed operations on given quantum states. In particular, we have focused on understanding the usefulness of pure entangled states of two modes (`fbits'), which are known to allow for Bell inequality violation only when at least two copies can be jointly processed~\cite{DasenbrookBowlesBohrBraskHoferFlindtBrunner2016}. Here, we find that single copies of such states can be useful for swapping the state of a single mode via teleportation. However, this procedure in itself is only useful (beyond classical notions of state transfer) when the latter mode is part of an entangled two-mode state itself. Once two fbits are available as a shared resource, one may teleport the entire two-mode state encoding the complex probability amplitudes usually encoded in a single qubit.

We have further considered how these teleportation protocols are influenced by particle-number superselection. Although not a fundamental restriction of Nature, it is of practical relevance for many applications, see, e.g., Ref.~\cite{OlofssonPottsBrunnerSamuelsson2020}. Here, we conclude that also this stronger superselection rule allows for teleportation with unit fidelity, when an appropriate four-mode resource state is shared, but using two fbits instead reduces the fidelity by  $50\%$, provided that no other practical restrictions limit the setup. Finally, we have discussed the peculiar notion of mixed maximally entangled states in the context of teleportation. Interestingly, such states are not merely an artefact of evaluating convex-roof entanglement measures under the restriction of superselection rules, but they do have limited usefulness for teleportation as well.

In comparison to usual qubit-based quantum information processing, fermionic systems hence provide a more differentiated picture of entanglement, non-locality, and teleportation. In this context, it may be of future interest to identify a suitably diverse set of entanglement quantifiers that can capture these different notions of useful entanglement. Such developments may further motivate revisiting previous observations about the energetic costs of creating correlations and entanglement~\cite{BruschiPerarnauLlobetFriisHovhannisyanHuber2015, FriisHuberPerarnauLlobet2016}. Moreover, one may even go as far as to speculate whether further differentiation of fermionic entanglement and correlations could become relevant to account for other applications, e.g., entanglement as a resource for fermionic measurement-based computation~\cite{ChiuChenChuang2013} or correlations relevant in molecular problems~\cite{DingSchilling2020}.


\vspace*{-2mm}
\begin{acknowledgments}
\vspace*{-1mm}
We are grateful to Nicolas Brunner, Edvin Olofsson, Patrick Potts, and Peter Samuelsson for fruitful discussions and valuable comments.
N.F acknowledges support from the Austrian Science Fund (FWF): P 31339-N27.
T.D. acknowledges support from the Brazilian agency CNPq INCT-IQ through the project (465469/2014-0). F.I. acknowledges the financial support of the Brazilian funding agencies CNPQ (Grant No.~308205/2019-7) and FAPERJ (Grant No.~E-26/211.318/2019).
GG acknowledges support by the Spanish \emph{Ministerio de Ciencia y Innovación} through the Project No. 2017-83780-P and by the EU through the FET-OPEN project SPRING ($\#$863098).\end{acknowledgments}

\bibliographystyle{apsrev4-1fixed_with_article_titles_full_names}
\bibliography{bibfile}

\end{document}